\documentclass[12pt,onecolumn,draftcls]{IEEEtran}
\usepackage[T1]{fontenc}
\usepackage[latin9]{inputenc}
\usepackage{color}
\usepackage{float}
\usepackage{amsmath}
\usepackage{amsthm}
\usepackage{amssymb}
\usepackage{graphicx}
\usepackage{setspace}
\usepackage[unicode=true,
 bookmarks=true,bookmarksnumbered=true,bookmarksopen=true,bookmarksopenlevel=1,
 breaklinks=false,pdfborder={0 0 0},pdfborderstyle={},backref=false,colorlinks=false]
 {hyperref}
\hypersetup{pdftitle={Your Title},
 pdfauthor={Your Name},
 pdfpagelayout=OneColumn, pdfnewwindow=true, pdfstartview=XYZ, plainpages=false}

\makeatletter

\providecommand{\tabularnewline}{\\}
\floatstyle{ruled}
\newfloat{algorithm}{tbp}{loa}
\providecommand{\algorithmname}{Algorithm}
\floatname{algorithm}{\protect\algorithmname}

\theoremstyle{plain}
\newtheorem{lem}{\protect\lemmaname}

\usepackage[caption=false,font=footnotesize]{subfig}
\usepackage{algorithmic}

\makeatother

\providecommand{\lemmaname}{Lemma}

\begin{document}
\title{\singlespacing{}Joint Pilot Optimization, Target Detection and Channel Estimation
for Integrated Sensing and Communication Systems\vspace{-0.2in}
}
\author{\singlespacing{}{\normalsize{}Zhe Huang, }\textit{\normalsize{}Student Member, IEEE}{\normalsize{},
Kexuan Wang, An Liu, }\textit{\normalsize{}Senior Member, IEEE}{\normalsize{},
Yunlong Cai,}\textit{\normalsize{} Senior Member, IEEE}{\normalsize{},
Rui Du and Tony Xiao Han}\thanks{This work was supported in part by National Science Foundation of
China (No.62071416), and in part by Huawei Technologies Co., Ltd.
(Corresponding authors: An Liu.)

Zhe Huang, Kexuan Wang, An Liu and Yunlong Cai are with the College
of Information Science and Electronic Engineering, Zhejiang University,
Hangzhou 310027, China (email: anliu@zju.edu.cn).

Rui Du and Tony Xiao Han are with Huawei Technologies Co., Ltd. (email:
tony.hanxiao@huawei.com).}}
\maketitle
\begin{abstract}
Radar sensing will be integrated into the 6G communication system
to support various applications. In this integrated sensing and communication
system, a radar target may also be a communication channel scatterer.
In this case, the radar and communication channels exhibit certain
joint burst sparsity. We propose a two-stage joint pilot optimization,
target detection and channel estimation scheme to exploit such joint
burst sparsity and pilot beamforming gain to enhance detection/estimation
performance. In Stage 1, the base station (BS) sends downlink pilots
(DP) for initial target search, and the user sends uplink pilots (UP)
for channel estimation. Then the BS performs joint target detection
and channel estimation based on the reflected DP and received UP signals.
In Stage 2, the BS exploits the prior information obtained in Stage
1 to optimize the DP signal to achieve beamforming gain and further
refine the performance. A Turbo Sparse Bayesian inference algorithm
is proposed for joint target detection and channel estimation in both
stages. The pilot optimization problem in Stage 2 is a semi-definite
programming with rank-1 constraints. By replacing the rank-1 constraint
with a tight and smooth approximation, we propose an efficient pilot
optimization algorithm based on the majorization-minimization method.
Simulations verify the advantages of the proposed scheme.
\end{abstract}


\section{Introduction}

It is expected that future 6G communication system will integrate
radar sensing and communication functions to support various important
application scenarios, such as autonomous drivingand smart cities
\cite{JRC}. Traditionally, radar sensing and communications are designed
separately as independent systems, and they usually occupy different
frequency bands to avoid interference. However, with the widespread
application of the massive multiple input multiple output (MIMO) and
millimeter wave (mmWave) communication technologies, future communication
signals will have higher time and angle resolution, which makes it
possible to use communication signals to achieve high-accuracy sensing.
Therefore, integrated sensing and communication (ISAC), in which the
radar sensing and communication sub-systems are jointly designed to
simultaneously achieve high-speed communication and high-accuracy
sensing using shared frequency band and hardware, has emerged as a
key technology in future communication systems \cite{JRC,JRC_2,IOR,Bliss}.
Recently, ISAC has attracted tremendous research interest in both
academia and industry \cite{SAR_com,Subcar_Power,OFDM_waveform,joint_coding_oppermann,Index_modulation}.
For example, there have been an increasing number of works on ISAC,
including the fundamental limits analysis \cite{zhang,bistatic_MAC},
transceiver architecture and frame structure \cite{JRC,Index_modulation},
ISAC waveform design \cite{OFDM_waveform,DMMSE_waveform,Range_sidelobe_waveform},
and temporal-spectral-spatial signal processing \cite{Subcar_Power,MU_JRC_beamforming,Bayes_JRC}.
These works show that there exist complex interplays between radar
sensing and communication. On the one hand, there is a tradeoff between
radar sensing and communication since they have to compete for the
same radio resource. On the other hand, radar sensing and communication
may help each other by providing useful side information to each other
and performing joint target detection and channel estimation.

\begin{figure}[tbh]
\begin{centering}
\includegraphics[width=120mm]{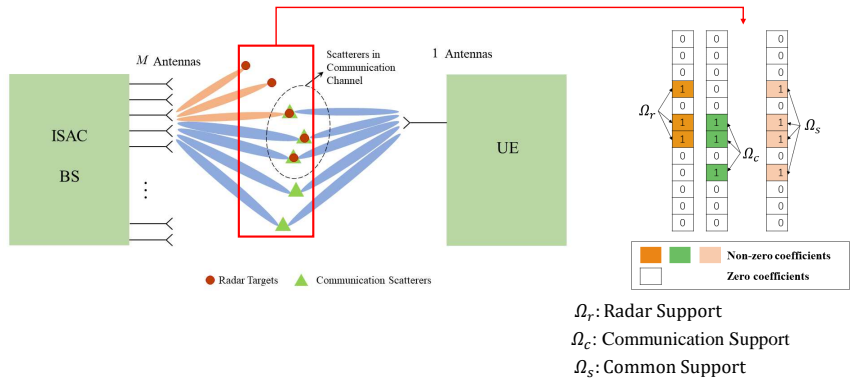}
\par\end{centering}
\centering{}\caption{\label{fig:Illustration-of-the}Illustration of the joint burst sparsity
in ISAC channels.}
\end{figure}

In this paper, we focus on an interesting interplay between radar
sensing and communication in massive MIMO ISAC system when the radar
and communication channels exhibit certain joint burst sparsity, as
illustrated in Fig. \ref{fig:Illustration-of-the}. Specifically,
in many cases, some radar targets are also communication scatterers.
As such, the angles of arrivals (AoAs) of the radar and communication
channels partially overlap. Moreover, both radar targets and communication
scatterers are usually concentrated in a few clusters, e.g., a large
target/scatterer can be viewed as a cluster of point targets/scatterers.
In this case, the AoAs of both radar and communication channels will
concentrate on a few non-zero bursts \cite{Joint_burst_LASSO}. Similar
correlations between the radar and communication channels have also
been reported in the literature. In \cite{JRC}, the communication
scatterers are assumed to be part of the radar targets, and thus the
AoAs of the communication channel is a subset of that of the radar
channel. In \cite{Caire_OTFSsensing}, each mobile user is treated
as a radar target and the AoA of the Line-of-Sight (LoS) path of the
communication channel is assumed to coincide with that of the radar
channel. In this case, the AoA obtained by the radar sensing can provide
partial channel state information (CSI) about the LoS path of the
communication channel, which can be exploited to design beamforming
for communications. The joint burst sparsity in this paper can be
viewed as a generalization of the correlation models for radar and
communication channels considered in \cite{JRC,Caire_OTFSsensing},
and is a more common situation in practical ISAC systems. Motivated
by the above observations, we propose to exploit the joint burst sparsity
of radar and communication channels for joint target detection and
channel estimation in massive MIMO ISAC system. Some related works
are summarized below.

\textbf{Massive MIMO channel estimation (CE):} Many works have been
devoted to address this problem \cite{Gao_TSP15_massiveMIMOCS,Rao_TSP15_massiveMIMOCS,Masood_TSP15_massiveMIMOCS}.
One major approach is to exploit the sparsity of massive MIMO channels
via compressive sensing (CS) to reduce the pilot overheads. In \cite{Masood_TSP15_massiveMIMOCS}
and \cite{rao2014distributed}, the joint sparsity between user antennas
that share some common scatterers has been exploited to design more
efficient compressive CE algorithms. In \cite{Liu_TWC2016_BurstLASSO},
the burst-sparse structure of massive MIMO channel has been exploited
to design a burst least absolute shrinkage and selection operator
(LASSO) CE algorithm. In \cite{Gao_TSP15_massiveMIMOCS,Rao_TSP15_massiveMIMOCS,Liu_TWC2016_BurstLASSO},
the temporal correlation of the channel support has been exploited
to reduce the CSI signaling overhead in massive MIMO systems. There
are also algorithms exploiting the joint burst sparsity of multi-user/multi-carrier
massive MIMO channels to further improve the CE performance, e.g.,
see \cite{Gao_TCOM2016_csMIMO}.

\textbf{Target detection and CE in massive MIMO ISAC System:}\textcolor{red}{{}
}Some recent works attempt to address target detection and CE in massive
MIMO ISAC system. In \cite{JRC}, the communication scatterers are
assumed to be part of the radar targets. However, the channel estimation
and target detection are performed separately based on the radar echo
signal and channel estimation pilots, respectively. In \cite{OTFS_joint},
the authors proposed to obtain the partial CSI about the LoS path
of the communication channel by using the BS as a radar to sense the
position of each mobile user. Specifically, they proposed a two-stage
target detection and CE scheme, in which the first stage performs
target detection, and the second stage performs super-resolution estimation
of the parameters associated with the radar target (i.e., the LoS
path parameters of the user channel).

\textbf{Pilot optimization for target detection:} A few works have
addressed the pilot optimization problem for target detection based
on Cramer-Rao Bound (CRB). In \cite{liu2021cramerrao}, a single target
detection problem has been considered and the pilot is optimized by
minimizing the trace of the Cramer-Rao Matrix based on semi-definite
relaxation (SDR). However, the SDR approach is only tight for some
special cases such as single target detection but in general suffers
from performance loss. In \cite{crb_eign}, the pilot has been optimized
by minimizing the maximum eigenvalue of the Cramer-Rao Matrix, also
through the SDR approach.

In the aforementioned studies, the radar target detection and communication
CE are performed separately based on the radar echo signal and CE
pilots, respectively. Moreover, it is very important to further enhance
the performance of both radar sensing and communication CE in the
low SNR regime in order to extend the coverage of ISAC systems, especially
for high-frequency band with larger path loss. However, how to achieve
high-accuracy radar sensing and CE in the low SNR regime remains a
challenging problem. In this paper, we propose a two-stage joint pilot
optimization, target detection and channel estimation (J-PoTdCe) scheme
to fully exploit the pilot beamforming gain and joint burst sparsity
of radar and communication channels for enhancing both the target
detection and channel estimation performance in massive MIMO ISAC
systems, especially for the low SNR regime. The main contributions
are summarized below.
\begin{itemize}
\item \textbf{Two-stage J-PoTdCe scheme:} We propose a two-stage J-PoTdCe
scheme so that the prior information obtained from Stage 1 can be
used to optimize the pilots and refine the detection/estimation performance
in Stage 2. Specifically, in Stage 1, the base station (BS) performs
joint target detection and channel estimation based on the reflected
omidirectional DP and received UP signals. In Stage 2, the BS exploits
the prior information obtained in Stage 1 to optimize the DP signal
to further refine the performance.
\item \textbf{Turbo-SBI algorithm:} We propose a hidden Markov model (HMM)
to capture the joint burst sparsity of the radar and communication
channels. Based on this model, a Turbo Sparse Bayesian inference (Turbo-SBI)
algorithm is proposed for\textbf{ }joint target detection and channel
estimation in both stages. Note that a Turbo-Orthogonal Approximate
Message Passing (OAMP) algorithm has been proposed in \cite{downlink_channel_estimation}
to exploit the joint burst sparsity of multi-user massive MIMO channels
under partially orthogonal (PO) measurement/pilot matrix. In this
paper, the associated measurement matrix is no longer PO because it
contains optimized pilot matrix and dynamic AoA grid parameters for
super-resolution AoA estimation. We show that the Turbo-OAMP can be
viewed as an approximation of the proposed Turbo-SBI for PO measurement
matrix.
\item \textbf{Pilot optimization based on rank-1 approximation and majorization-minimization
(MM):} The pilot optimization problem in Stage 2 is formulated as
a semi-definite programming with rank-1 constraints, which aims at
exploring the beamforming gain and minimizing the worst-case Cramer-Rao
Bound (CRB) of the target parameters. By replacing the rank-1 constraint
with a tight and smooth approximation, we propose an efficient pilot
optimization algorithm based on the MM method. Compared with the conventional
SDR algorithm in \cite{liu2021cramerrao,crb_eign}, the proposed pilot
optimization algorithm has similar complexity order but better performance
since it directly takes into account the rank 1 constraints in the
algorithm design.
\end{itemize}
Finally, the advantages of the proposed J-PoTdCe scheme and the associated
Turbo-SBI and pilot optimization algorithms are verified by simulations
under the clustered delay line (CDL) channel model in 3GPP R15 \cite{3gpp_Rel15}.
The rest of the paper is organized as follows. In Section II, we describe
the system model and the overall two-stage J-PoTdCe scheme. In Section
III, we present the Turbo-SBI algorithm for\textbf{ }joint target
detection and channel estimation in both stages. In Section IV, we
present the CRB analysis and the pilot optimization algorithm in Stage
2. The simulation results and conclusions are given in Section V and
VI, respectively.

\section{Two-Stage J-PoTdCe Scheme}

In this section, we describe the system model and the proposed two-stage
J-PoTdCe scheme. Consider a TDD massive MIMO ISAC system with one
BS serving a single-antenna mobile user while detecting $K$ targets
indexed by $k\in\left\{ 1,\ldots,K\right\} $, as illustrated in Fig.
\ref{fig:Illustration-of-the}. The BS is equipped with $M\gg1$ antennas.
In the channel estimation phase, we will focus on one single-antenna
user for clarity. However, the proposed J-PoTdCe scheme can be readily
extended to the case with multiple multi-antenna users, by assigning
orthogonal uplink pilots (UPs) for different antennas. While all targets
reflect back the echo wave to the BS, not all of them contribute to
communication paths between the BS and the user \cite{JRC}. Therefore,
it is natural to assume that there is a partial overlap between $K$
targets and $L$ communication scatterers. Note that we do not explicitly
add clutters in the system model due to the following reasons. On
one hand, the effects of weak clutters can be absorbed into the noise.
On the other hand, the strong clutters can be treated as targets of
non-interest, whose parameters will also be estimated to mitigate
the interference caused by strong clutters and enhance the detection
performance of the targets of interest. After detecting all the targets
(interest or non-interest), the targets of interest can be further
identified by exploiting the properties/features of their parameters.
For ISAC systems, detecting the strong clutters may also help enhancing
the channel estimation performance of the user because some strong
clutters may also contribute to the communication paths.

\subsection{Outline of the Two-Stage J-PoTdCe}

\begin{figure}[tbh]
\begin{centering}
\includegraphics[width=110mm]{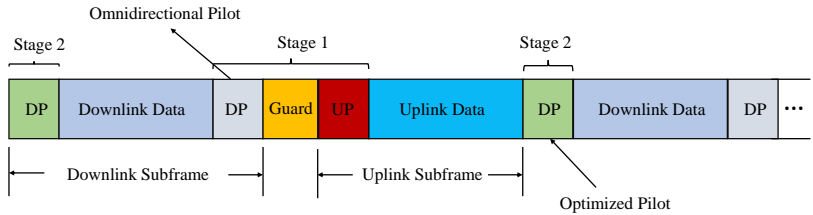}
\par\end{centering}
\centering{}\caption{\label{fig:The-two-stage-J-PoTdCe}Frame structure of the two-stage
J-PoTdCe scheme.}
\end{figure}

In the two-stage J-PoTdCe scheme, the time axis is divided into frames,
and each frame contains two phases: target detection and channel estimation
phase and data transmission phase. In this paper, we will focus on
the first phase, which can be further divided into the following two
stages as shown in\textcolor{red}{{} }Fig. \ref{fig:The-two-stage-J-PoTdCe}:
\begin{itemize}
\item Initial target detection and channel estimation (Stage 1): Stage 1
is to search for potential targets, and provide an initial estimation
for the target parameters and communication channels. After Stage
1, the BS will have some prior information about the target/channel
parameters, e.g., whether there is a target or communication scatterer
in a certain direction. Such prior information can be exploited to
optimize the pilots in the second stage. Specifically, the BS first
sends $P_{1}$ omnidirectional DPs for initial target search. Then
the user sends $Q$ UPs to the BS for channel estimation. Finally,
the BS performs the joint target detection and channel estimation
based on the reflected DP and received UP signals.
\item Refined target detection and channel estimation (Stage 2): Based on
the prior information about the targets and channel obtained in the
initial stage, the BS optimizes the pilots and sends $P_{2}$ directional
DPs towards the targets and communication scatterers for more accurate
observations. Finally, the BS refines the joint target detection and
channel estimation based on the reflected DP signals in both stages
as well as the UP signals in Stage 1.
\end{itemize}
Note that in the above descriptions, we have ignored the data transmissions
for conciseness. In the frame structure in Fig. \ref{fig:The-two-stage-J-PoTdCe},
the omnidirectional DPs in Stage 1 are actually transmitted at the
end of the downlink subframe. Then the UPs in Stage 1 are transmitted
at the beginning of the uplink subframe followed by the uplink data
transmission. Finally, the DPs in Stage 2 are transmitted at the beginning
of the next downlink subframe. Therefore, the channel and target parameters
are assumed to be (approximately) constant with the duration of one
subframe.

\subsection{Reflected DP Signal Model for Target Detection}

In the $p$-th DP symbol duration of Stage $t$ ($t\in1,2$), the
BS transmits a DP $\boldsymbol{v}_{t,p}\in\mathbb{C}^{M}$, and the
corresponding received signal can be expressed as
\begin{equation}
\boldsymbol{y}_{t,p}^{r}=\mathbf{H}^{r}\boldsymbol{v}_{t,p}+\boldsymbol{n}_{t,p}^{r},\label{eq:downlink_signal}
\end{equation}
where $\mathbf{H}^{r}\in\mathbb{C}^{M\times M}$ is the radar channel
matrix and $\boldsymbol{n}_{t,p}^{r}\sim\mathcal{CN}\left(0,\left(\sigma_{n}^{r}\right)^{2}\mathbf{I}\right)\in\mathbb{C}^{M}$
is the additive white Gaussian noise (AWGN) with variance $\left(\sigma_{n}^{r}\right)^{2}$.
For convenience, define the aggregate received DP signal (radar measurements)
of all the $P_{t}$ pilot symbols as $\boldsymbol{y}_{t}^{r}\triangleq\left[(\boldsymbol{y}_{t,1}^{r})^{T},...,(\boldsymbol{y}_{t,P_{t}}^{r})^{T}\right]^{T}\in\mathbb{C}^{P_{t}M\times1}$.
The radar channel matrix depends on the AoAs and radar cross sections
(RCSs) of the targets and can be modeled as
\begin{equation}
\mathbf{H}^{r}=\sum_{k=1}^{K}x_{k}^{r}\boldsymbol{a}\left(\theta_{k}^{r}\right)\boldsymbol{a}^{H}\left(\theta_{k}^{r}\right),\label{eq:radarchannel}
\end{equation}
where $\theta_{k}^{r}$ and $x_{k}^{r}$ are the AoA and RCS of the
$k$-th target, $\boldsymbol{a}\left(\theta\right)\in\mathbb{C}^{M}$
is the array response vector for the BS antenna array. For a half-wavelength
space uniform linear array (ULA), the array response vector is given
by

\[
\boldsymbol{a}\left(\theta\right)=\frac{1}{\sqrt{M}}\left[1,e^{-j\pi\sin\theta},\ldots,e^{-j\left(M-1\right)\pi\sin\theta}\right]^{T}.
\]

\subsection{Received UP Signal Model for Channel Estimation}

In the $q$-th UP symbol duration of Stage 1, the user transmits an
uplink pilot $u_{q}\in\mathbb{C}$ and the corresponding received
signal can be expressed as
\begin{equation}
\boldsymbol{y}_{q}^{c}=\mathbf{h}^{c}u_{q}+\boldsymbol{n}_{q}^{c},\label{eq:uplink_signal}
\end{equation}
where $\mathbf{h}^{c}\in\mathbb{C}^{M}$ is the communication channel
vector and $\boldsymbol{n}_{q}^{c}\sim\mathcal{CN}\left(0,\left(\sigma_{n}^{c}\right)^{2}\mathbf{I}\right)\in\mathbb{C}^{M}$
is the AWGN. For convenience, define the aggregate received UP signal
(channel measurements) of all the $Q$ pilot symbols as $\boldsymbol{y}^{c}\triangleq\left[(\boldsymbol{y}_{1}^{c})^{T},...,(\boldsymbol{y}_{Q}^{c})^{T}\right]\in\mathbb{C}^{QM\times1}$.
The communication channel vector can be modeled as
\begin{equation}
\mathbf{h}^{c}=\sum_{l=1}^{L}x_{l}^{c}\boldsymbol{a}\left(\theta_{l}^{c}\right),\label{eq:comchannel}
\end{equation}
where $\theta_{l}^{c}$ and $x_{l}^{c}$ are the AoA and complex gain
of the $l$-th channel path, respectively.

Note that for clarity, we focus on a narrowband ISAC system with low-speed
targets and users in this paper. In a wideband ISAC system with range
and/or Doppler estimation capability, the model in (\ref{eq:radarchannel})
and (\ref{eq:comchannel}) should also include the range/delay and
Doppler of the targets/channel paths. Typically, in these ISAC systems,
the estimation for the direction (AoA), range and Doppler of the targets/channel
paths is implemented by processing the receiving channels over time
and obtaining multi-channel measurements for each considered range-Doppler
bin \cite{Compress_sensing_technologies_for_next_generation_wireless_communication}.
The model in (\ref{eq:radarchannel}) and (\ref{eq:comchannel}) refers
to a single range-Doppler bin \cite{Compress_sensing_technologies_for_next_generation_wireless_communication}.
Therefore, the joint target detection and channel estimation algorithm
in this paper can be applied to detect/estimate multiple targets/channel
paths for each range-Doppler bin in a wideband ISAC system.

To complete the proposed two-stage J-PoTdCe scheme, we need to design
the joint target detection and channel estimation algorithm for both
stages, as well as the pilot optimization algorithm in Stage 2, which
will be elaborated in Section \ref{sec:Joint-Target-Detection} and
\ref{sec:Optimal-Pilot-Design}, respectively.

\section{Joint Target Detection and Channel Estimation Algorithm\label{sec:Joint-Target-Detection}}

\subsection{Sparse Angular Domain Channel with Dynamic Grid}

We first describe the sparse angular domain representation for the
radar and communication channels, which is a necessary step in order
to apply the sparse recovery methods such as sparse Bayesian inference.
One commonly used method to obtain a sparse representation of the
channel is to define a uniform grid $\left\{ \overline{\theta}_{1},...,\overline{\theta}_{\widetilde{M}}\right\} $
of $\widetilde{M}\gg K+L$ AoA points, such that $[sin\overline{\theta}_{1},...,sin\overline{\theta}_{\widetilde{M}}]$
is uniformly spaced over $[-1,1]$. If the AoAs of the targets and
channel paths indeed take values in the discrete set $\left\{ \overline{\theta}_{1},...,\overline{\theta}_{\widetilde{M}}\right\} $,
the radar and communication channels in (\ref{eq:radarchannel}) and
(\ref{eq:comchannel}) can be rewritten as
\begin{align}
\mathbf{H}^{r} & =\mathbf{A}\text{Diag}(\boldsymbol{x}^{r})\mathbf{A}^{H}=\sum_{m=1}^{\widetilde{M}}x_{m}^{r}\boldsymbol{a}\left(\overline{\theta}_{m}\right)\boldsymbol{a}^{H}\left(\overline{\theta}_{m}\right),\label{eq:angelchannelr}\\
\mathbf{h}^{c} & =\mathbf{A}\boldsymbol{x}^{c}=\sum_{m=1}^{\widetilde{M}}x_{m}^{c}\boldsymbol{a}\left(\overline{\theta}_{m}\right),\label{eq:angelchannelc}
\end{align}
where $\mathbf{A}\triangleq\left[\boldsymbol{a}\left(\overline{\theta}_{1}\right),\cdots,\boldsymbol{a}\left(\overline{\theta}_{\widetilde{M}}\right)\right]$
is a fixed array response matrix corresponding to the uniform grid,
$x_{m}^{r}$ is the radar cross section(RCS) of the target in the
$m$-th AoA direction $\overline{\theta}_{m}$, and $x_{m}^{c}$ is
the complex gain of the channel path from the user to the $m$-th
AoA direction $\overline{\theta}_{m}$ at the BS. For convenience,
we define $\boldsymbol{x}^{r}\triangleq\left[x_{1}^{r},...,x_{\widetilde{M}}^{r}\right]^{T}\in\mathbb{C}^{\widetilde{M}}$
as the angular domain radar channel, and $\boldsymbol{x}^{c}\triangleq\left[x_{1}^{c},...,x_{\widetilde{M}}^{c}\right]^{T}\in\mathbb{C}^{\widetilde{M}}$
as the angular domain communication channel. If there is no target
(active channel path) in the $m$-th AoA direction, we have $x_{m}^{r}=0$
($x_{m}^{c}=0$). Therefore, there are $K$ ($L$) non-zero elements
in $\boldsymbol{x}^{r}$ ($\boldsymbol{x}^{c}$) corresponding to
the $K$ targets ($L$ active channel paths). Note that, we use $x_{m}^{r}$
and $x_{m}^{c}$ to denote the RCS of the target and complex channel
gain in the $m$-th AoA direction, respectively, even though $x_{k}^{r}$
and $x_{l}^{c}$ have been used to denote the RCS of the $k$-th radar
target and complex gain of the $l$-th active channel path in Section
II.

In practice, however, the true AoA may not lie exactly on the $\widetilde{M}$
discrete AoA grid points. As a result, we need to use a very large
$\widetilde{M}$ in order to achieve a high AoA estimation accuracy,
leading to a high computational complexity. To overcome the above
mismatch and complexity issues of using a fixed grid, we adopt dynamic
grid parameters $\boldsymbol{\theta}\triangleq\left[\theta_{1},...,\theta_{\widetilde{M}}\right]^{T}$.
In this case, as long as $\widetilde{M}\geq K+L$, there always exist
a set of unknown (and potentially non-uniform) grid parameters $\boldsymbol{\theta}$
that can exactly represent the true radar and communication channels
by
\begin{align*}
\mathbf{H}^{r} & =\mathbf{A}(\boldsymbol{\theta})\text{Diag}(\boldsymbol{x}^{r})\mathbf{A}(\boldsymbol{\theta})^{H},\\
\mathbf{h}^{c} & =\mathbf{A}(\boldsymbol{\theta})\boldsymbol{x}^{c},
\end{align*}
where $\mathbf{A}(\boldsymbol{\theta})\triangleq\left[\boldsymbol{a}\left(\theta_{1}\right),\cdots,\boldsymbol{a}\left(\theta_{\widetilde{M}}\right)\right]$.
However, if we set $\widetilde{M}=K+L$ exactly, the likelihood function
associated with the estimation of the dynamic grids $\boldsymbol{\theta}$
will have many local maxima, making it difficult to obtain an accurate
estimation of $\boldsymbol{\theta}$ using the maximum likelihood
(ML) method, as the algorithm can easily get stuck in a ``bad''
local maxima. If $\widetilde{M}$ is sufficiently large, then by using
a uniform grid as the initial point for $\boldsymbol{\theta}$, each
true AoA will be very close to one initial grid point, making it much
easier for the algorithm to find a near-optimal solution for the ML
estimation problem. In the rest of the paper, we set $\widetilde{M}=M$
to achieve a good tradeoff between the AoA estimation performance
and complexity, since the AoA resolution for a massive MIMO array
with $M\gg1$ is roughly $O\left(\frac{1}{M}\right)$.

One may argue that when $\widetilde{M}=M$, the total number of radar
and channel measurements $M(P_{1}+P_{2}+Q)$ is no less than the total
number of parameters $3M$, and thus there is no need to use sparse
recovery methods. However, a properly designed sparse recovery algorithm
can fully exploit the joint burst sparsity to mitigate the noise effect
and significantly enhance the overall performance in the low SNR regime,
as will be shown in the simulations.

\subsection{Hidden Markov Model for Joint Burst Sparsity}

In practice, the radar and communication channels exhibit certain
joint burst sparsity as explained in the introduction and illustrated
in Fig. \ref{fig:Illustration-of-the}. In this section, we shall
introduce a hidden Markov model to capture the joint burst sparse
structure of the radar and communication channels. Specifically, let
$\boldsymbol{s}^{r}=\left[s_{1}^{r},...,s_{M}^{r}\right]^{T}$ and
$\mathbf{\mathit{\boldsymbol{s}}}^{c}=\left[s_{1}^{c},...,s_{M}^{c}\right]^{T}$
denote the support vectors of the radar and communication channels,
respectively, where $s_{m}^{r}=1$ ($s_{m}^{c}=1$) indicates there
is a radar target (communication scatterer) around the $m$-th AoA
grid $\theta_{m}$, and $s_{m}^{r}=0$ ($s_{m}^{c}=0$) indicates
the opposite. Therefore, in Fig. \ref{fig:Illustration-of-the}, $\Omega_{r}\triangleq\left\{ m:s_{m}^{r}=1\right\} $
indicates the set of (coarse) AoAs of radar targets, $\Omega_{c}\triangleq\left\{ m:s_{m}^{c}=1\right\} $
indicates the set of (coarse) AoAs of user, and $\Omega_{s}\triangleq\Omega_{r}\bigcup\Omega_{c}$
indicates the common AoA set.

Conditioned on the channel support vectors $\boldsymbol{s}^{r}$ and
$\boldsymbol{s}^{c}$, the elements of $\boldsymbol{x}^{r}$ and $\boldsymbol{x}^{c}$
are independent and the conditional prior distributions are respectively
given by
\begin{equation}
p(x_{m}^{r}|s_{m}^{r})=(1-s_{m}^{r})\delta(x_{m}^{r})+s_{m}^{r}\mathcal{CN}\left(x_{m}^{r};0,(\sigma_{m}^{r})^{2}\right)
\end{equation}
\begin{equation}
p(x_{m}^{c}|s_{m}^{c})=(1-s_{m}^{c})\delta(x_{m}^{c})+s_{m}^{c}\mathcal{CN}\left(x_{m}^{c};0,(\sigma_{m}^{c})^{2}\right),
\end{equation}
where $(\sigma_{m}^{r})^{2}$ and $(\sigma_{m}^{c})^{2}$ are the
variance of $x_{m}^{r}$ and $x_{m}^{c}$ conditioned on $s_{m}^{r}=1$
and $s_{m}^{c}=1$, respectively.

To represent the common AoAs of the radar and communication channels,
a joint support vector $\boldsymbol{s}=\left[s_{1},...,s_{M}\right]\in\left\{ 0,1\right\} ^{M}$
with $s_{m}=s_{m}^{r}\lor s_{m}^{c}$ is introduced in the HMM, where
$\lor$ represents the logical ``or'' operator. The joint distribution
for the channel support vectors $\boldsymbol{s}$, $\boldsymbol{s}^{r}$
and $\boldsymbol{s}^{c}$ is given by $p(\boldsymbol{s},\boldsymbol{s}^{r},\boldsymbol{s}^{c})=p(\boldsymbol{s})p(\boldsymbol{s}^{r}|\boldsymbol{s})p(\boldsymbol{s}^{c}|\boldsymbol{s})$,
where

\begin{equation}
p(\boldsymbol{s}^{r}|\boldsymbol{s})=\prod_{m}\mathop{p(s_{m}^{r}|s_{m})}=\prod_{m}(1-s_{m})\delta(s_{m}^{r})+s_{m}\rho_{r}^{s_{m}^{r}}(1-\rho_{r})^{1-s_{m}^{r}},
\end{equation}

\begin{equation}
p(\boldsymbol{s}^{c}|\boldsymbol{s})=\prod_{m}\mathop{p(s_{m}^{c}|s_{m})}=\prod_{m}(1-s_{m})\delta(s_{m}^{c})+s_{m}\rho_{c}^{s_{m}^{c}}(1-\rho_{c})^{1-s_{m}^{c}}
\end{equation}

where $\rho_{r}=\frac{\left|\Omega_{r}\right|}{\left|\Omega_{s}\right|}$
$(\rho_{c}=\frac{\left|\Omega_{c}\right|}{\left|\Omega_{s}\right|})$
is the probability of $s_{m}^{r}=1$ ($s_{m}^{c}=1$) conditioned
on $s_{m}=1$, which measures the degree of overlapping between the
targets and communication scatterers. Furthermore, to capture the
burst sparse structure of the joint communication and radar channel,
the joint support vector $\boldsymbol{s}$ is modeled as a Markov
chain:
\begin{equation}
p(\boldsymbol{s})=p\left(s_{1}\right)\prod_{m=1}^{M-1}\mathop{p(s_{m+1}|s_{m})},\label{eq:sMarkov-1}
\end{equation}
with the transition probability given by $p(s_{m+1}=1|s_{m}=0)=\rho_{0,1}$
and $p\left(s_{m+1}=0|s_{m}=1\right)=\rho_{1,0}$.\textcolor{black}{{}
The initial distribution $p\left(s_{1}\right)$ is set to be the steady
state distribution of the Markov chain in (\ref{eq:sMarkov-1}), i.e.,
\begin{equation}
\lambda\triangleq p\left(s_{m}=1\right)=\frac{\rho_{0,1}}{\rho_{0,1}+\rho_{1,0}}.\label{eq:deflamR-1}
\end{equation}
The transition probabilities }$\rho_{0,1}$ and $\rho_{1,0}$ determine
the average length of each non-zero burst and the total number of
non-zero bursts in $\boldsymbol{s}$, and $\lambda$ determines the
sparsity level of $\boldsymbol{s}$.

Finally, the joint prior distribution of all the random variables
in HMM is given by

\begin{equation}
p(\boldsymbol{s},\boldsymbol{s}^{r},\boldsymbol{s}^{c},\boldsymbol{x}^{r},\boldsymbol{x}^{c})=p(\boldsymbol{s})\prod_{m}\mathop{p(s_{m}^{r}|s_{m})}\prod_{m}\mathop{p(s_{m}^{c}|s_{m})}\prod_{m}p(x_{m}^{r}|s_{m}^{r})p(x_{m}^{c}|s_{m}^{c}).
\end{equation}

\subsection{Sparse Bayesian Inference Formulation for Joint Detection and Estimation}

The problem formulation and algorithm design for the two stages can
be unified by using the same notation $\boldsymbol{v}_{1},...,\boldsymbol{v}_{P}$
and $\boldsymbol{y}_{1}^{r},...,\boldsymbol{y}_{P}^{r}$ to denote
the DPs and the received DP signals in both stages. Specifically,
in Stage 1, we have $P=P_{1}$ and $\boldsymbol{v}_{p}=\boldsymbol{v}_{1,p},\boldsymbol{y}_{p}^{r}=\boldsymbol{y}_{1,p}^{r},p=1,...,P_{1}$.
In Stage 2, we have $P=P_{1}+P_{2}$, $\boldsymbol{v}_{p}=\boldsymbol{v}_{1,p},\boldsymbol{y}_{p}^{r}=\boldsymbol{y}_{1,p}^{r},p=1,...,P_{1}$
and $\boldsymbol{v}_{P_{1}+p}=\boldsymbol{v}_{2,p},\boldsymbol{y}_{P_{1}+p}^{r}=\boldsymbol{y}_{2,p}^{r},p=1,...,P_{2}$.
For convenience, we define the radar and communication measurement
matrices $\mathbf{F}^{r}\text{(\ensuremath{\boldsymbol{\theta}})}\triangleq\mathbf{V}\mathbf{\tilde{A}}(\boldsymbol{\theta})\in\mathbb{C}^{PM\times M}$
and $\mathbf{F}^{c}\text{(\ensuremath{\boldsymbol{\theta}})}\triangleq\mathbf{U}\mathbf{A}\text{(\ensuremath{\boldsymbol{\theta}})}\text{\ensuremath{\in\mathbb{C}^{QM\times M}}}$,
where
\[
\mathbf{V}=\left[\begin{array}{c}
\boldsymbol{v}_{1}^{T}\otimes\mathbf{I}_{M}\\
\ldots\\
\boldsymbol{v}_{P}^{T}\otimes\mathbf{I}_{M}
\end{array}\right],\text{ }\mathbf{U}=\left[\begin{array}{c}
u_{1}\mathbf{I}_{M}\\
\ldots\\
u_{Q}\mathbf{I}_{M}
\end{array}\right],
\]
$\mathbf{\tilde{A}}(\boldsymbol{\theta})\in\mathbb{C}^{M^{2}\times M}$
consists of the $\left(m-1\right)M+m$-th column of $\mathbf{A}^{*}(\boldsymbol{\theta})\otimes\mathbf{A}(\boldsymbol{\theta})$
for $m=1,....,M$, and $\otimes$ means the Kronecker product. Using
these notations, (\ref{eq:downlink_signal}) and (\ref{eq:uplink_signal})
can be rewritten as a linear observation model as
\begin{equation}
\boldsymbol{y}=\mathbf{F}\text{(\ensuremath{\boldsymbol{\theta}})}\boldsymbol{x}+\boldsymbol{n},
\end{equation}
where $\boldsymbol{y}^{c}=\left[(\boldsymbol{y}^{r})^{T},(\boldsymbol{y}^{c})^{T}\right]^{T}$,
$\boldsymbol{y}^{r}=\left[(\boldsymbol{y}_{1}^{r})^{T},...,(\boldsymbol{y}_{P}^{r})^{T}\right]^{T}$,
$\boldsymbol{x}=\left[(\boldsymbol{x}^{r})^{T},(\boldsymbol{x}^{c})^{T}\right]^{T}$,
$\boldsymbol{n}$ is the aggregated noise vector and $\mathbf{F}(\boldsymbol{\theta})=\text{BlockDiag}\left(\mathbf{F}^{r}(\boldsymbol{\theta}),\mathbf{F}^{c}(\boldsymbol{\theta})\right)$.

\textcolor{black}{For given gird parameter $\boldsymbol{\theta}$
and observation $\boldsymbol{y}$, we aim at computing the conditional
marginal posteriors $p\left(\boldsymbol{x}^{r}|\boldsymbol{y},\boldsymbol{\theta}\right)$,
$p\left(\boldsymbol{x}^{c}|\boldsymbol{y},\boldsymbol{\theta}\right)$,
$p\left(s_{m}^{r}|\boldsymbol{y},\boldsymbol{\theta}\right),\forall m$
(i.e., perform Bayesian inference for $\boldsymbol{x}^{r},\boldsymbol{x}^{c}$
and $s_{m}^{r},s_{m}^{c},\forall m$). On the other hand, the grid
parameter is obtained by ML estimation as 
\begin{align}
\boldsymbol{\theta}^{*} & =\underset{\boldsymbol{\theta}}{\text{argmax}}\ln p(\boldsymbol{y}|\boldsymbol{\theta}).\label{eq:ML_Para}
\end{align}
Once we have the ML estimate of $\boldsymbol{\theta}$ and the associated
conditional marginal posteriors, the MAP estimate of the communication
channel as $\boldsymbol{x}^{c*}=\text{argmax}_{\boldsymbol{x}^{c}}p\left(\boldsymbol{x}^{c}|\boldsymbol{y},\boldsymbol{\theta}^{*}\right)$
and $\boldsymbol{h}^{c*}=\mathbf{A}(\boldsymbol{\theta}^{*})\boldsymbol{x}^{c*}$
can be obtained. Moreover, $p\left(s_{m}^{r}|\boldsymbol{y},\boldsymbol{\theta}^{*}\right)$
gives the probability that a target exists at the AoA direction $\theta_{m}^{*}$.}

\textcolor{black}{It is very challenging to calculate the above conditional
marginal posteriors because the factor graph of the underlying probability
model has loops. In the following subsections, we shall propose a
Turbo-SBI algorithm which approximately calculates the marginal posteriors
and finds an approximate solution for (\ref{eq:ML_Para}). The proposed
Turbo-SBI algorithm is shown in the simulations to achieve a good
performance.}

\subsection{Outline of the Turbo-SBI Algorithm}

\textcolor{black}{Based on the Expectation-maximization (EM) method,
the Turbo-SBI algorithm starts from the uniform grid $\boldsymbol{\theta}^{0}$
and performs iterations between the following two major steps until
convergence. }
\begin{itemize}
\item \textbf{\textcolor{black}{Turbo-SBI-E Step:}}\textcolor{black}{{} For
given grid parameter $\boldsymbol{\theta}^{i}$ in the $i$-th iteration,
we approximately calculate the posteriors $p\left(\boldsymbol{x}^{r}|\boldsymbol{y},\boldsymbol{\theta}^{i}\right)$,
$p\left(\boldsymbol{x}^{c}|\boldsymbol{y},\boldsymbol{\theta}^{i}\right)$,
$p\left(s_{m}^{r}|\boldsymbol{y},\boldsymbol{\theta}^{i}\right),\forall m$
by combining the message passing and LMMSE approaches via the turbo
framework;}
\item \textbf{\textcolor{black}{Turbo-SBI-M Step:}}\textcolor{black}{{} Using
the approximate posterior $p\left(\boldsymbol{x}|\boldsymbol{y},\boldsymbol{\theta}^{i}\right)$
obtained in the }Turbo-SBI-E\textcolor{black}{{} Step, calculate the
gradient for the likelihood function $\ln p(\boldsymbol{y}|\boldsymbol{\theta})$
at $\boldsymbol{\theta}^{i}$, then use gradient ascent update to
obtain the next iterate $\boldsymbol{\theta}^{i+1}$.}
\end{itemize}
\textcolor{black}{}

\textcolor{black}{In the following two subsections, we first elaborate
how to approximately calculate the posterior $p\left(\boldsymbol{x}|\boldsymbol{y},\boldsymbol{\theta}^{i}\right)$
and the other marginal posteriors $p\left(s_{m}^{r}|\boldsymbol{y},\boldsymbol{\theta}^{i}\right),\forall m$
in the Turbo-VBI-E Step. Then we present the Turbo-VBI-M Step, which
requires the posterior $p\left(\boldsymbol{x}|\boldsymbol{y},\boldsymbol{\theta}^{i}\right)$
calculated in the }Turbo-SBI-E\textcolor{black}{{} step. Note that similar
to the grid parameter $\boldsymbol{\theta}$, the parameters $\rho_{r},\rho_{c},\rho_{0,1},\rho_{1,0}$
in the HMM prior model can also be automatically learned based on
the EM method. Please refer to \cite{vila2011expectation} for the
details of the EM method to learn the parameters in the prior model.}

\subsection{Turbo-SBI-E Step with Given Grid Parameters}

\begin{figure}[tbh]
\begin{centering}
\includegraphics[width=100mm]{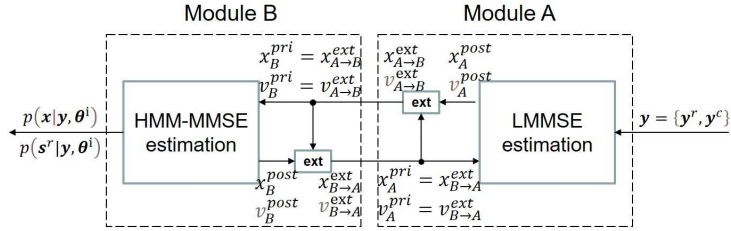}
\par\end{centering}
\centering{}\caption{\label{fig:Turbo-SBI}Modules of the Turbo-SBI-E step and message
flow between different modules.}
\end{figure}

The Turbo-SBI-E Step contains two modules, as shown in Fig. \ref{fig:Turbo-SBI}.
Module A is a LMMSE estimator based on the observation $\boldsymbol{y}$
and extrinsic messages from Module B. Module B, which is called the
HMM-MMSE estimator, performs MMSE estimation that combines the HMM
prior and the extrinsic messages from Module A. The two modules are
executed iteratively until convergence. In the following, we elaborate
the two modules in Fig. \ref{fig:Turbo-SBI}. Since the grid is fixed
in the Turbo-SBI-E Step, we shall omit the grid $\boldsymbol{\theta}$
in $\mathbf{F}\text{(\ensuremath{\boldsymbol{\theta}})}$ in this
subsection.

\subsubsection*{A. LMMSE in Module A}

In Module A, we assume that $\boldsymbol{x}$ follows a Gaussian distribution
with a prior mean $\boldsymbol{x}_{A}^{pri}=\boldsymbol{x}_{B\rightarrow A}^{ext}$
and covariance $\boldsymbol{V}_{A}^{pri}=\boldsymbol{V}_{B\rightarrow A}^{ext}$,
where $\boldsymbol{x}_{B\rightarrow A}^{ext}$ and $\boldsymbol{V}_{B\rightarrow A}^{ext}$
are the extrinsic message output from Module B, as will be given in
(\ref{eq:extBtoA}). Note that $\boldsymbol{V}_{B\rightarrow A}^{ext}$
is a diagonal matrix. With this assumption and the linear observation
model $\boldsymbol{y}=\mathbf{F}\boldsymbol{x}+\boldsymbol{n},$ the
posterior mean of $\boldsymbol{x}$ is given by the LMMSE estimator
\begin{equation}
\boldsymbol{x}_{A}^{post}=\boldsymbol{V}_{A}^{post}\left((\boldsymbol{V}_{A}^{pri})^{-1}\boldsymbol{x}_{A}^{pri}+\frac{\mathbf{F}^{H}\boldsymbol{y}}{\sigma_{n}^{2}}\right)\label{eq:xApost}
\end{equation}
and $\boldsymbol{V}_{A}^{post}$ is the posterior covariance of $\boldsymbol{x}$
given by
\begin{equation}
\boldsymbol{V}_{A}^{post}=\left(\frac{\mathbf{F}^{H}\mathbf{F}}{\sigma_{n}^{2}}+(\boldsymbol{V}_{A}^{pri})^{-1}\right)^{-1}.\label{eq:VApost}
\end{equation}
In the simulations, we note that the off-diagonal elements of $\boldsymbol{V}_{A}^{inv}\triangleq\frac{\mathbf{F}^{H}\mathbf{F}}{\sigma_{n}^{2}}+(\boldsymbol{V}_{A}^{pri})^{-1}$
are usually much smaller than its diagonal elements, and most non-zero
off-diagonal elements concentrate on the five-diagonal sub-matrix
of $\boldsymbol{V}_{A}^{inv}$. In fact, for uniform grid $\boldsymbol{\theta}$,
$\boldsymbol{V}_{A}^{inv}$ reduces to a diagonal matrix. Let $\boldsymbol{V}_{A,0}^{inv}$
denote the five-diagonal sub-matrix of $\boldsymbol{V}_{A}^{inv}$.
By applying the first-order Taylor expansion of $(\boldsymbol{V}_{A}^{inv})^{-1}$
at $\boldsymbol{V}_{A,0}^{inv}$, the calculation of $\boldsymbol{V}_{A}^{post}$
can be safely approximated as
\begin{equation}
\boldsymbol{V}_{A}^{post}\approx2\boldsymbol{V}_{A,0}^{inv}-(\boldsymbol{V}_{A,0}^{inv})^{-1}\boldsymbol{V}_{A}^{inv}(\boldsymbol{V}_{A,0}^{inv})^{-1}.\label{eq:Vinv}
\end{equation}
Then the extrinsic message passed to Module B can be calculated by
excluding the prior information $\boldsymbol{x}_{A}^{pri},\boldsymbol{V}_{A}^{pri}$
as
\begin{align}
\boldsymbol{V}_{A\rightarrow B}^{ext} & =\left((\overline{\boldsymbol{V}}_{A}^{post})^{-1}-(\boldsymbol{V}_{A}^{pri})^{-1}\right)^{-1},\nonumber \\
\boldsymbol{x}_{A\rightarrow B}^{ext} & =\boldsymbol{V}_{A\rightarrow B}^{ext}\left((\overline{\boldsymbol{V}}_{A}^{post})^{-1}\boldsymbol{x}_{A}^{post}-(\boldsymbol{V}_{A}^{pri})^{-1}\boldsymbol{x}_{A}^{pri}\right),\label{eq:extAtoB}
\end{align}
where $\overline{\boldsymbol{V}}_{A}^{post}$ is a diagonal approximation
of $\boldsymbol{V}_{A}^{post}$ by setting the off-diagonal elements
to zeros. The above approximations in (\ref{eq:Vinv}) and $\overline{\boldsymbol{V}}_{A}^{post}$
can greatly simplify the calculations with little performance loss,
as verified by simulations.

\subsubsection*{B. Message Passing in Module B}

\begin{figure}[tbh]
\centering{}\includegraphics[width=70mm]{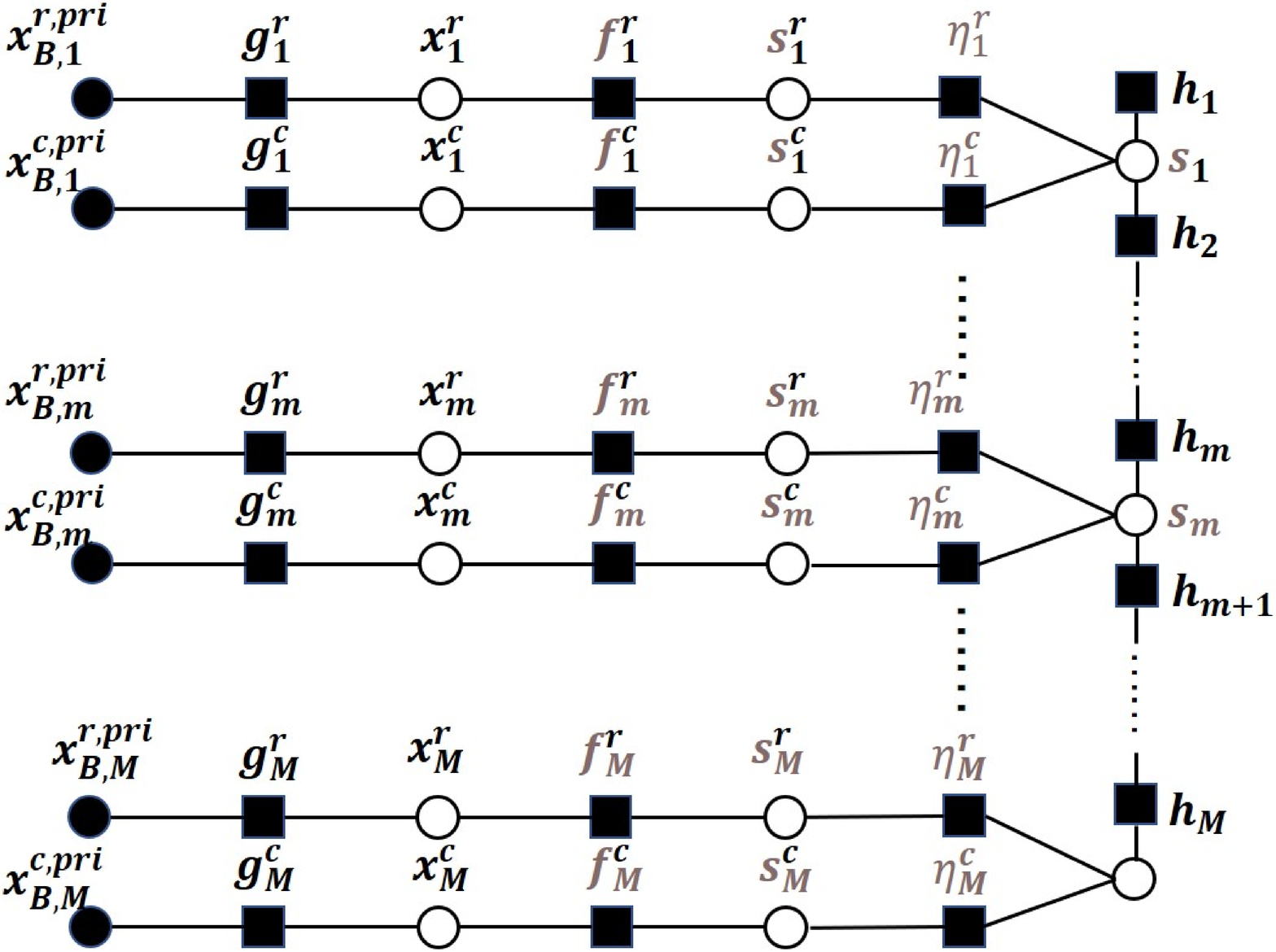}\caption{\label{fig:Factor-graph}Factor graph of the joint distribution $p(\boldsymbol{s},\boldsymbol{s}^{r},\boldsymbol{s}^{c},\boldsymbol{x}^{r},\boldsymbol{x}^{c},\boldsymbol{x}_{B}^{r,pri},\boldsymbol{x}_{B}^{c,pri}|\boldsymbol{\theta})$.}
\end{figure}

In Module B, a message passing scheme is used for the HMM-MMSE estimator
to calculate the posterior of $\boldsymbol{x}$ and $\boldsymbol{s}^{r}$,
based on the HMM channel prior and the extrinsic messages $\boldsymbol{x}_{A\rightarrow B}^{ext},\boldsymbol{V}_{A\rightarrow B}^{ext}$
from Module A. Specifically, the extrinsic messages $\boldsymbol{x}_{A\rightarrow B}^{ext},\boldsymbol{V}_{A\rightarrow B}^{ext}$
are equivalently modeled as a virtual AWGN observation model:
\[
\boldsymbol{x}_{B}^{r,pri}=\boldsymbol{x}^{r}+\mathbf{z}^{r},
\]
\[
\boldsymbol{x}_{B}^{c,pri}=\boldsymbol{x}^{c}+\mathbf{z}^{c},
\]
where the extrinsic mean $\boldsymbol{x}_{A\rightarrow B}^{ext}=[(\boldsymbol{x}_{B}^{r,pri})^{T},(\boldsymbol{x}_{B}^{c,pri})^{T}]^{T}$
is treated as observations obtained via a virtual AWGN channel with
zero mean noise vectors $\mathbf{z}^{r}$ and $\mathbf{z}^{c}$, and
the extrinsic covariance $\boldsymbol{V}_{A\rightarrow B}^{ext}=\text{BlockDiag}\left(\boldsymbol{V}_{B}^{r,pri},\boldsymbol{V}_{B}^{c,pri}\right)$
is treated as the noise covariance, i.e., $\mathbf{z}^{r}\boldsymbol{\sim}\mathcal{CN}\left(0;\boldsymbol{V}_{B}^{r,pri}\right)$,
$\mathbf{z}^{c}\boldsymbol{\sim}\mathcal{CN}\left(0;\boldsymbol{V}_{B}^{c,pri}\right)$.
Similar treatment has been used in various approximate message passing
algorithms, see e.g., \cite{Bayati_TIT11_AMPSEprrof,Schniter_TSP12_TurboAMP,ma2015turbo}
for justifications of this treatment. The factor graph $\mathcal{G}_{B}$
of the joint distribution associated with this virtual AWGN observation
model is shown in Fig. \ref{fig:Factor-graph}, where the function
expression of each factor node is listed in Table \ref{tab:Factor-Distri-func}.
In Table\textcolor{red}{{} }\ref{tab:Factor-Distri-func}, $x_{B,m}^{r,pri}$
and $x_{B,m}^{c,pri}$ are the $m$-th elements of $\boldsymbol{x}_{B}^{r,pri}$and
$\boldsymbol{x}_{B}^{c,pri}$, respectively, and $v_{B,m}^{r,pri}$
and $v_{B,m}^{c,pri}$ are the $m$-th diagonal elements of $\boldsymbol{V}_{B}^{r,pri}$and
$\boldsymbol{V}_{B}^{r,pri}$, respectively.

\begin{table*}[tbh]
\begin{centering}
\caption{\label{tab:Factor-Distri-func}Factors, distributions and functional
forms in Fig. \ref{fig:Factor-graph}.}
\par\end{centering}
\centering{}%
\begin{tabular}{|c|c|c|}
\hline 
Factor & Distribution & Functional form\tabularnewline
\hline 
\hline 
{\small{}$\begin{array}{c}
g_{m}^{r}\left(x_{B,m}^{r,pri},x_{m}^{r}\right)\\
g_{m}^{c}\left(x_{B,m}^{c,pri},x_{m}^{c}\right)
\end{array}$} & {\small{}$\begin{array}{c}
p\left(x_{m}^{r}|x_{B,m}^{r,pri}\right)\\
p\left(x_{m}^{c}|x_{B,m}^{c,pri}\right)
\end{array}$} & {\small{}$\begin{array}{c}
\mathcal{CN}\left(x_{m}^{r};x_{B,m}^{r,pri},v_{B,m}^{r,pri}\right)\\
\mathcal{CN}\left(x_{m}^{c};x_{B,m}^{c,pri},v_{B,m}^{c,pri}\right)
\end{array}$}\tabularnewline
\hline 
\hline 
{\small{}$\begin{array}{c}
f_{m}^{r}\left(x_{m}^{r},s_{m}^{r}\right)\\
f_{m}^{c}\left(x_{m}^{c},s_{m}^{c}\right)
\end{array}$} & {\small{}$\begin{array}{c}
p\left(x_{m}^{r}|s_{m}^{r}\right)\\
p\left(x_{m}^{c}|s_{m}^{c}\right)
\end{array}$} & {\small{}$\begin{array}{c}
\left(1-s_{m}^{r}\right)\delta\left(x_{m}^{r}\right)+s_{m}^{r}\mathcal{CN}\left(x_{m}^{r};0,\left(\sigma_{m}^{r}\right)^{2}\right)\\
\left(1-s_{m}^{c}\right)\delta\left(x_{m}^{c}\right)+s_{m}^{c}\mathcal{CN}\left(x_{m}^{c};0,\left(\sigma_{m}^{c}\right)^{2}\right)
\end{array}$}\tabularnewline
\hline 
\hline 
{\small{}$\begin{array}{c}
\eta_{m}^{r}\left(s_{m}^{r},s_{m}\right)\\
\eta_{m}^{c}\left(s_{m}^{c},s_{m}\right)
\end{array}$} & {\small{}$\begin{array}{c}
p\left(s_{m}^{r}|s_{m}\right)\\
p\left(s_{m}^{c}|s_{m}\right)
\end{array}$} & {\small{}$\begin{array}{c}
p\left(s_{m}^{r}=1|s_{m}=0\right)=0,p\left(s_{m}^{r}=1|s_{m}=1\right)=\rho^{r}\\
p\left(s_{m}^{c}=1|s_{m}=0\right)=0,p\left(s_{m}^{c}=1|s_{m}=1\right)=\rho^{c}
\end{array}$}\tabularnewline
\hline 
\hline 
{\small{}$h_{1}^{s}(s_{1})$} & {\small{}$p\left(s_{1}\right)$} & {\small{}$\left(\lambda\right)^{s_{1}}\left(1-\lambda\right)^{1-s_{1}}$}\tabularnewline
\hline 
\hline 
{\small{}$h_{m+1}^{s}\left(s_{m+1},s_{m}\right)$} & {\small{}$p\left(s_{m+1}|s_{m}\right)$} & {\small{}$\begin{cases}
\left(\rho_{0,1}\right)^{s_{m+1}}\left(1-\rho_{0,1}\right)^{1-s_{m+1}}, & s_{m}=0\\
\left(1-\rho_{1,0}\right)^{s_{m+1}}\left(\rho_{1,0}\right)^{1-s_{m+1}}, & s_{m}=1
\end{cases}$}\tabularnewline
\hline 
\end{tabular}
\end{table*}

We now outline the message passing scheme over the factor graph $\mathcal{G}_{B}$.
The details are elaborated in Appendix \ref{subsec:Message-Update-Equations}.
According to the sum-product rule, the message passing over $x_{m}^{r}\rightarrow f_{m}^{r}\rightarrow s_{m}^{r}\rightarrow\eta_{m}^{r}\rightarrow s_{m}$
and $x_{m}^{c}\rightarrow f_{m}^{c}\rightarrow s_{m}^{c}\rightarrow\eta_{m}^{c}\rightarrow s_{m}$
are given by (\ref{eq:message1}) - (\ref{eq:message4}). Then a forward
backward message passing is performed over the Markov chains $\boldsymbol{s}$
through (\ref{eq:message5}) - (\ref{eq:message8}). After this, the
message is passed back over the path $s_{m}\rightarrow\eta_{m}^{r}\rightarrow s_{m}^{r}\rightarrow f_{m}^{r}\rightarrow x_{m}^{r}$
and $s_{m}\rightarrow\eta_{m}^{c}\rightarrow s_{m}^{c}\rightarrow f_{m}^{c}\rightarrow x_{m}^{c}$
using (\ref{eq:message9}) - (\ref{eq:message11}).

After calculating the updated messages \{$v_{\boldsymbol{f}_{m}^{r}\rightarrow\boldsymbol{x}_{m}^{r}}$\},
the approximate posterior distributions are given by 
\begin{equation}
\hat{p}(x_{m}^{r}|\boldsymbol{y})\propto v_{\boldsymbol{f}_{m}^{r}\rightarrow\boldsymbol{x}_{m}^{r}}\times v_{\boldsymbol{x}_{m}^{r}\rightarrow\boldsymbol{f}_{m}^{r}},\label{eq:xpostB}
\end{equation}
\begin{equation}
\hat{p}\left(s_{m}^{r}|\boldsymbol{y}\right)=\frac{\pi_{s^{r},m}^{in}\pi_{s^{r},m}^{out}}{\pi_{s^{r},m}^{in}\pi_{s^{r},m}^{out}+(1-\pi_{s^{r},m}^{in})(1-\pi_{s^{r},m}^{out})},\forall m,\label{eq:spostB}
\end{equation}
where $v_{\boldsymbol{f}_{m}^{r}\rightarrow\boldsymbol{x}_{m}^{r}}$,
$v_{\boldsymbol{x}_{m}^{r}\rightarrow\boldsymbol{f}_{m}^{r}}$, $\pi_{s^{r},m}^{in}$,
$\pi_{s^{r},m}^{out}$ are given in Appendix \ref{subsec:Message-Update-Equations}.
Then the posterior mean $\boldsymbol{x}_{B}^{r,post}=\left[x_{B,1}^{r,post},...,x_{B,M}^{r,post}\right]^{T}$
and variance $\boldsymbol{V}_{B}^{r,post}=\text{Diag}([v_{B,1}^{post},...,v_{B,M}^{post}])$
for $\boldsymbol{x}^{r}$ can be respectively calculated as
\begin{equation}
x_{B,m}^{r,post}=\int_{x_{m}^{r}}x_{m}^{r}\hat{p}(x_{m}^{r}|\boldsymbol{y}),\label{eq:xmeanpostB}
\end{equation}
\begin{equation}
v_{B,m}^{post}=\int_{x_{m}^{r}}|x_{m}-x_{B,m}^{r,post}|^{2}\hat{p}(x_{m}^{r}|\boldsymbol{y}),\label{eq:vmeanpostb}
\end{equation}
for $m=1,...,M$. The posterior mean $\boldsymbol{x}_{B}^{c,post}$
and variance $\boldsymbol{V}_{B}^{c,post}$ for $\boldsymbol{x}^{c}$
can be calculated similarly. Then the extrinsic message passed to
Module A can be calculated as
\begin{align}
\boldsymbol{V}_{B\rightarrow A}^{ext} & =\left((\boldsymbol{V}_{B}^{post})^{-1}-(\boldsymbol{V}_{B}^{pri})^{-1}\right)^{-1},\nonumber \\
\boldsymbol{x}_{B\rightarrow A}^{ext} & =\boldsymbol{V}_{B\rightarrow A}^{ext}\left((\boldsymbol{V}_{B}^{post})^{-1}\boldsymbol{x}_{B}^{post}-(\boldsymbol{V}_{B}^{pri})^{-1}\boldsymbol{x}_{B}^{pri}\right),\label{eq:extBtoA}
\end{align}
where $\boldsymbol{x}_{B}^{pri}=[(\boldsymbol{x}_{B}^{r,pri})^{T},(\boldsymbol{x}_{B}^{c,pri})^{T}]^{T}$,
$\boldsymbol{x}_{B}^{post}=[(\boldsymbol{x}_{B}^{r,post})^{T},(\boldsymbol{x}_{B}^{c,post})^{T}]^{T}$,
$\text{\ensuremath{\boldsymbol{V}_{B}^{pri}}}\triangleq\textrm{BlockDiag}\left(\boldsymbol{V}_{B}^{r,pri},\boldsymbol{V}_{B}^{c,pri}\right)$
and $\text{\ensuremath{\boldsymbol{V}_{B}^{post}\triangleq}BlockDiag}\left(\boldsymbol{V}_{B}^{r,post},\boldsymbol{V}_{B}^{c,post}\right)$.

Finally, we point out that the Turbo-OAMP in \cite{downlink_channel_estimation}
is an approximation of the proposed Turbo-SBI-E Step when $\mathbf{F}$
is PO. Specifically, in Turbo-OAMP, by assuming a PO measurement matrix
$\mathbf{F}$, the $\mathbf{F}^{H}\mathbf{F}$ in the LMMSE Step in
(\ref{eq:VApost}) is approximated as $\mathbf{F}^{H}\mathbf{F}\approx\frac{\text{tr}\left(\mathbf{F}^{H}\mathbf{F}\right)}{2M}\boldsymbol{I}$.
Moreover, when calculating the extrinsic messages in (\ref{eq:extAtoB})
and (\ref{eq:extBtoA}), $\boldsymbol{V}_{B}^{post}$ and $\boldsymbol{V}_{B}^{post}$
are approximated as $v_{A}^{post}\boldsymbol{I}$ and $v_{B}^{post}\boldsymbol{I}$,
respectively, where $v_{A}^{post}$ and $v_{B}^{post}$ are the mean
values of the diagonal elements of $\boldsymbol{V}_{B}^{post}$ and
$\boldsymbol{V}_{B}^{post}$, respectively.

\subsection{Turbo-SBI-M Step}

In the M step, we need to maximize \textcolor{black}{the log-likelihood
function $\ln p(\boldsymbol{y}|\boldsymbol{\theta})$, which is difficult
because $\ln p(\boldsymbol{y}|\boldsymbol{\theta})$ does not have
a closed-form expression. Inspired by the EM method, we construct
a surrogate function for $\ln p(\boldsymbol{y}|\boldsymbol{\theta})$
around the current iterate $\boldsymbol{\theta}^{i}$ as follows}
\begin{align*}
Q(\boldsymbol{\theta};\boldsymbol{\theta}^{i}) & =\int p(\boldsymbol{x}|\boldsymbol{y},\boldsymbol{\theta}^{i})\mathrm{ln}\frac{p(\boldsymbol{y},\boldsymbol{x}|\boldsymbol{\theta})}{p(\boldsymbol{x}|\boldsymbol{y},\boldsymbol{\theta}^{i})}d\boldsymbol{x}\\
 & =-\frac{||\boldsymbol{y}-\mathbf{F}(\boldsymbol{\theta})\boldsymbol{x}^{post}||_{2}^{2}+\text{tr}(\mathbf{F}(\boldsymbol{\theta})\boldsymbol{V}^{post}\mathbf{F}(\boldsymbol{\theta})^{H})}{\sigma_{n}^{2}}+c,
\end{align*}
where $\boldsymbol{x}^{post}$ and $\boldsymbol{V}^{post}$ denote
the posterior mean and covariance associated with $p(\boldsymbol{x}|\boldsymbol{y},\boldsymbol{\theta}^{i})$,
and $c$ is a constant. It can be shown that $Q(\boldsymbol{\theta};\boldsymbol{\theta}^{i})\leq\ln p(\boldsymbol{y}|\boldsymbol{\theta}),\forall\boldsymbol{\theta}$,
$Q(\boldsymbol{\theta}^{i};\boldsymbol{\theta}^{i})=\ln p(\boldsymbol{y}|\boldsymbol{\theta}^{i})$
and $\nabla_{\boldsymbol{\theta}}Q(\boldsymbol{\theta}^{i};\boldsymbol{\theta}^{i})=\nabla_{\boldsymbol{\theta}}\ln p(\boldsymbol{y}|\boldsymbol{\theta}^{i})$.
Based on this, the next iterate $\boldsymbol{\theta}^{i+1}$ can be
obtained using the gradient ascent method as
\begin{equation}
\boldsymbol{\theta}^{i+1}=\boldsymbol{\theta}^{i}+\tau^{i}\nabla_{\boldsymbol{\theta}}Q(\boldsymbol{\theta}^{i};\boldsymbol{\theta}^{i}),\label{eq:thetaGradup}
\end{equation}
where $\tau^{i}$ is the step size which can be determined by applying
the Arjimo rule to $Q(\boldsymbol{\theta};\boldsymbol{\theta}^{i})$.
The Arjimo rule ensures that $Q(\boldsymbol{\theta}^{i+1};\boldsymbol{\theta}^{i})\geq Q(\boldsymbol{\theta}^{i};\boldsymbol{\theta}^{i})$
and the equality only holds when $\boldsymbol{\theta}^{i}$ is already
a stationary point of the ML estimation problem. Therefore, we have
$\ln p(\boldsymbol{y}|\boldsymbol{\theta}^{i+1})\geq Q(\boldsymbol{\theta}^{i+1};\boldsymbol{\theta}^{i})\geq Q(\boldsymbol{\theta}^{i};\boldsymbol{\theta}^{i})=\ln p(\boldsymbol{y}|\boldsymbol{\theta}^{i})$,
i.e., the Turbo-SBI-M Step can strictly increase the likelihood function
until convergence to a stationary point. Finally, the gradient $\nabla_{\boldsymbol{\theta}}Q(\boldsymbol{\theta}^{i};\boldsymbol{\theta}^{i})=\left[\frac{\partial Q(\boldsymbol{\theta}^{i};\boldsymbol{\theta}^{i})}{\partial\theta_{1}},...,\frac{\partial Q(\boldsymbol{\theta}^{i};\boldsymbol{\theta}^{i})}{\partial\theta_{M}}\right]^{T}$
is given by
\begin{align}
\frac{\partial Q(\boldsymbol{\theta}^{i};\boldsymbol{\theta}^{i})}{\partial\theta_{m}}= & 2\mathrm{Re}[\boldsymbol{a}^{\prime}(\theta_{m}^{i})^{H}\mathbf{U}^{H}\mathbf{U}\boldsymbol{a}(\theta_{m}^{i})c_{1}^{i}+\boldsymbol{a}^{\prime}(\theta_{m}^{i})^{H}\mathbf{U}^{H}\boldsymbol{c}_{2}^{i}]\nonumber \\
+ & 2\mathrm{Re}[\tilde{\boldsymbol{a}}^{\prime}(\theta_{m}^{i})^{H}\mathbf{V}^{H}\mathbf{V}\tilde{\boldsymbol{a}}(\theta_{m}^{i})c_{3}^{i}+\tilde{\boldsymbol{a}}^{\prime}(\theta_{m}^{i})^{H}\mathbf{V}^{H}\boldsymbol{c}_{4}^{i}],\label{eq:Qgrad}
\end{align}
where $\tilde{\boldsymbol{a}}(\theta_{m}^{i})$ is the $m$-th column
of $\mathbf{\tilde{A}}(\boldsymbol{\theta})$, $\boldsymbol{y}_{-m}^{c,i}=\boldsymbol{y}^{c}-\mathbf{U}\underset{j\neq m}{\sum}\left(x_{A,j}^{c,post}\cdot\boldsymbol{a}\left(\theta_{j}^{i}\right)\right),$

$\boldsymbol{y}_{-m}^{r,i}=\boldsymbol{y}^{r}-\mathbf{\mathbf{\mathbf{V}}}\underset{j\neq m}{\sum}\left(x_{A,j}^{r,post}\cdot\tilde{\boldsymbol{a}}\left(\theta_{j}^{i}\right)\right),$
$\boldsymbol{a}^{\prime}\left(\theta_{m}^{i}\right)=d\boldsymbol{a}\left(\boldsymbol{\theta}^{i}\right)/d\theta_{m}^{i},$
$\tilde{\boldsymbol{a}}_{m}^{\prime}\left(\theta_{m}^{i}\right)=d\tilde{\boldsymbol{a}}_{m}(\boldsymbol{\theta}^{i})/d\theta_{m}^{i},$

$c_{1}^{i}=-\sigma_{n}^{-2}\left(\left|x_{A,m}^{c,post}\right|^{2}+\mathbf{\mathrm{\mathit{v}}}_{A,m}^{c,post}\right)$,
$\boldsymbol{c}_{2}^{i}=\sigma_{n}^{-2}\left(\left(x_{A,m}^{c,post}\right)^{\ast}\boldsymbol{y}_{-m}^{c,i}-\mathbf{U}\underset{j\neq m}{\sum}\mathbf{\mathrm{\mathit{v}}}_{A,j}^{r,post}\boldsymbol{a}\left(\theta_{j}^{i}\right)\right)$,

$c_{3}^{i}=-\sigma_{n}^{-2}\left(\left|x_{A,m}^{r,post}\right|^{2}+\mathbf{\mathrm{\mathit{v}}}_{A,m}^{r,post}\right)$,
$\boldsymbol{c}_{4}^{i}=\sigma_{n}^{-2}\left(\left(x_{A,m}^{r,post}\right)^{\ast}\boldsymbol{y}_{-m}^{r,i}-\mathbf{V}\underset{j\neq m}{\sum}\mathbf{\mathrm{\mathit{v}}}_{A,j}^{r,post}\tilde{\boldsymbol{a}}\left(\theta_{j}^{i}\right)\right)$.

Note that to calculate $Q(\boldsymbol{\theta};\boldsymbol{\theta}^{i})$
and its gradient, we need to know the posterior mean and covariance
$\boldsymbol{x}^{post}$ and $\boldsymbol{V}^{post}$ associated with
$p(\boldsymbol{x}|\boldsymbol{y},\boldsymbol{\theta}^{i})$, which
can be approximated using the $\boldsymbol{x}_{B}^{post}$ and $\boldsymbol{V}_{B}^{post}$
calculated in the Turbo-SBI-E Step. Finally, the overall Turbo-SBI
algorithm is summarized in Algorithm \ref{alg1}.

\begin{algorithm}[tbh]
{\small{}\caption{\label{alg1}Turbo-SBI algorithm}
}{\small\par}

\textbf{Input:} $\boldsymbol{y}$, $\boldsymbol{\theta}^{0}$, \textcolor{black}{maximum
iteration numbers $I_{in},I_{out}$, threshold $\epsilon$.}

\textbf{Output:} $\boldsymbol{\theta}^{\ast},$ $\boldsymbol{x}^{*},$
$\hat{p}\left(s_{m}^{r}|\boldsymbol{y},\boldsymbol{\theta}^{*}\right),\forall m$.

\begin{algorithmic}

\FOR{${\color{blue}{\color{black}i=1,\cdots,I_{out}}}$}

\STATE \textbf{Turbo-SBI-E Step:}

\STATE Initialize $i_{in}=1$, $\boldsymbol{x}_{A}^{pri}=\boldsymbol{0}$
and $\boldsymbol{V}_{A}^{pri}$.

\WHILE{not converge and $i_{in}\leq I_{in}$}

\STATE $i_{in}=i_{in}+1$.

\STATE \textbf{\%Module A: LMMSE Estimator}

\STATE Update $\boldsymbol{x}_{A}^{post}$ and $V_{A}^{post}$, using
(\ref{eq:xApost}) and (\ref{eq:VApost})/(\ref{eq:Vinv}).

\STATE Update $\boldsymbol{x}_{B}^{pri}=\boldsymbol{x}_{A\rightarrow B}^{ext}$
and $\boldsymbol{V}_{B}^{pri}=\boldsymbol{V}_{A\rightarrow B}^{ext}$,
using (\ref{eq:extAtoB}).

\STATE\textbf{\%Module B: HMM-MMSE Estimator}

\STATE Perform message passing over the factor graph $\mathcal{G}_{B}$
using (\ref{eq:message1}) - (\ref{eq:message11}).

\STATE Calculate the approximate posterior distributions $\hat{p}(x_{m}^{r}|\boldsymbol{y},\boldsymbol{\theta}^{i}),\hat{p}\left(s_{m}^{r}|\boldsymbol{y},\boldsymbol{\theta}^{i}\right),\forall m$
using (\ref{eq:xpostB}), (\ref{eq:spostB}).

\STATE Update $\boldsymbol{x}_{B}^{post}$ and $\ensuremath{\boldsymbol{V}_{B}^{post}}$
using (\ref{eq:xmeanpostB}) and (\ref{eq:vmeanpostb}).

\STATE Update $\boldsymbol{x}_{A}^{pri}=\boldsymbol{x}_{B\rightarrow A}^{ext}$
and $\boldsymbol{V}_{A}^{pri}=\boldsymbol{V}_{B\rightarrow A}^{ext}$,
using (\ref{eq:extBtoA}).

\ENDWHILE

\STATE \textbf{Turbo-SBI-M Step:}

\STATE Calculate the gradient $\frac{\partial Q(\boldsymbol{\theta}^{i};\boldsymbol{\theta}^{i})}{\partial\theta_{m}}$
in (\ref{eq:Qgrad}) using the $\boldsymbol{x}_{B}^{post}$ and $\ensuremath{\boldsymbol{V}_{B}^{post}}$
from the E step.

\STATE Obtain $\boldsymbol{\theta}^{i+1}$ using the gradient ascent
update in (\ref{eq:thetaGradup}).

\IF{\textcolor{black}{$\left\Vert \boldsymbol{\theta}^{i+1}-\boldsymbol{\theta}^{i}\right\Vert \leq\epsilon$}}

\STATE \textbf{\textcolor{black}{break}}

\ENDIF

\ENDFOR

\STATE Output $\boldsymbol{\theta}^{\ast}$, $\boldsymbol{x}^{*}=\boldsymbol{x}_{B}^{post}$
and $\hat{p}\left(s_{m}^{r}|\boldsymbol{y},\boldsymbol{\theta}^{*}\right)$.

\end{algorithmic}
\end{algorithm}

\subsection{\textcolor{black}{Complexity Analysis of Turbo-SBI}}

The complexity of Module A is mainly dominated by the matrix inverse
operation in (\ref{eq:VApost}), whose complexity is $O\left(M^{3}\right)$.
By using the first-order Taylor expansion in (\ref{eq:Vinv}), we
can reduce the complexity of Module A to $O\left(M^{2}\right)$. The
complexity of Module B is $O(M)$ since it only involves scalar or
diagonal matrix operations. Finally, the complexity of the gradient{\small{}
}ascent update for the off-grid parameter in (\ref{eq:thetaGradup})
is dominated by the matrix multiplication $\mathbf{V}\tilde{\boldsymbol{a}}(\theta_{m}^{i})$,
whose complexity is $O\left(M^{2}P\right)$. Therefore, the overall
per outer iteration complexity of the Turbo-SBI is $O\left(I_{in}M^{2}+M^{2}P\right)$.

\section{Optimal Pilot Design based on Cramer-Rao Bound\label{sec:Optimal-Pilot-Design}}

In this section, we present the pilot design based on the estimated
AoAs in Stage 1. We first derive the Cramer-Rao Bound (CRB) of the
AoAs. Then we formulate the pilot optimization problem as a worst-case
CRB minimization problem. Finally, we propose an efficient algorithm
to solve the pilot optimization problem.

\subsection{Derivation of Cramer-Rao Bound}

In this subsection, we derive the CRB of the AoAs under the assumption
of known channel coefficients $x_{k}^{r}$'s and $x_{l}^{c}$'s. The
unknown AoA parameters $\boldsymbol{\theta}=\left[\begin{array}{ccc}
\left(\boldsymbol{\theta}^{r}\right)^{T}, & \left(\boldsymbol{\theta}^{s}\right)^{T}, & \left(\boldsymbol{\theta}^{c}\right)^{T}\end{array}\right]^{T}$ are divided into three subsets, namely, the AoAs of the purely radar
targets $\boldsymbol{\theta}^{r}$, the AoAs of the purely communication
paths $\boldsymbol{\theta}^{c}$, and the common AoAs $\boldsymbol{\theta}^{s}$.
Note that, we use $\boldsymbol{\theta}$ to denote the set of all
AoA parameters in this section, even though $\boldsymbol{\theta}$
has been used to denote the dynamic grid in Section III. As in \cite{kay1993fundamentals},
the Fisher Information Matrix (FIM) $\mathbf{J\left(\boldsymbol{\theta}\right)}$
is defined by

\begin{equation}
\mathbf{J\left(\boldsymbol{\theta}\right)=\mathbb{E}}\left\{ \left[\frac{\partial\ln p\left(\boldsymbol{y}\mid\boldsymbol{\theta}\right)}{\partial\boldsymbol{\theta}}\right]\left[\frac{\partial\ln p\left(\boldsymbol{y}\mid\boldsymbol{\theta}\right)}{\partial\boldsymbol{\theta}}\right]^{T}\right\} ,
\end{equation}
where $p\left(\boldsymbol{y}\mid\boldsymbol{\theta}\right)$ is the
likelihood function of the observation $\boldsymbol{y}$ and $\frac{\partial\ln p\left(\boldsymbol{y}\mid\boldsymbol{\theta}\right)}{\partial\boldsymbol{\theta}}$
is the gradient vector of the log-likelihood function with respect
to $\boldsymbol{\theta}$. According to this definition, the FIM based
on the reflected DP signals and received UP signals is given by

\begin{equation}
\mathbf{J}\left(\boldsymbol{\theta}\right)=\left[\begin{array}{ccc}
\mathbf{J}\left(\boldsymbol{\theta}^{r},\boldsymbol{\theta}^{r}\right) & \mathbf{J}\left(\boldsymbol{\theta}^{r},\boldsymbol{\theta}^{s}\right) & \mathbf{J}\left(\boldsymbol{\theta}^{r},\boldsymbol{\theta}^{c}\right)\\
\mathbf{J}\left(\boldsymbol{\theta}^{r},\boldsymbol{\theta}^{s}\right)^{T} & \mathbf{J}\left(\boldsymbol{\theta}^{s},\boldsymbol{\theta}^{s}\right) & \mathbf{J}\left(\boldsymbol{\theta}^{s},\boldsymbol{\theta}^{c}\right)\\
\mathbf{J}\left(\boldsymbol{\theta}^{r},\boldsymbol{\theta}^{c}\right)^{T} & \mathbf{J}\left(\boldsymbol{\theta}^{s},\boldsymbol{\theta}^{c}\right)^{T} & \mathbf{J}\left(\boldsymbol{\theta}^{c},\boldsymbol{\theta}^{c}\right)
\end{array}\right],
\end{equation}
and the submatrices in $\mathbf{J}\left(\boldsymbol{\theta}\right)$
are given by
\begin{equation}
\mathbf{J}\left(\boldsymbol{\theta}^{r},\boldsymbol{\theta}^{c}\right)=\mathbf{0},\mathbf{J}\left(\boldsymbol{\theta}^{c},\boldsymbol{\theta}^{c}\right)=2\left(\sigma_{n}^{c}\right)^{-2}\mathrm{Re}\left\{ \frac{\partial\boldsymbol{h}^{c}}{\partial\boldsymbol{\theta}^{c}}\left(\boldsymbol{\mathbf{\Psi}}^{c}\right)^{T}\left(\boldsymbol{\Psi}^{c}\right)^{*}\left(\frac{\partial\boldsymbol{h}^{c}}{\partial\boldsymbol{\theta}^{c}}\right)^{H}\right\} 
\end{equation}
\begin{equation}
\mathbf{J}\left(\boldsymbol{\theta}^{r},\boldsymbol{\theta}^{r}\right)=2\left(\sigma_{n}^{r}\right)^{-2}\mathrm{Re}\left\{ \frac{\partial\boldsymbol{h}^{r}}{\partial\boldsymbol{\theta}^{r}}\left(\boldsymbol{\mathbf{\Psi}}_{1}^{r}\right)^{T}\left(\boldsymbol{\mathbf{\Psi}}_{1}^{r}\right)^{*}\left(\frac{\partial\boldsymbol{h}^{r}}{\partial\boldsymbol{\theta}^{r}}\right)^{H}\right\} +2\left(\sigma_{n}^{r}\right)^{-2}\mathrm{Re}\left\{ \frac{\partial\boldsymbol{h}^{r}}{\partial\boldsymbol{\theta}^{r}}\left(\boldsymbol{\mathbf{\Psi}}_{2}^{r}\right)^{T}\left(\boldsymbol{\mathbf{\Psi}}_{2}^{r}\right)^{*}\left(\frac{\partial\boldsymbol{h}^{r}}{\partial\boldsymbol{\theta}^{r}}\right)^{H}\right\} ,
\end{equation}
\begin{equation}
\mathbf{J}\left(\boldsymbol{\theta}^{r},\boldsymbol{\theta}^{s}\right)=2\left(\sigma_{n}^{r}\right)^{-2}\mathrm{Re}\left\{ \frac{\partial\boldsymbol{h}^{r}}{\partial\boldsymbol{\theta}^{r}}\left(\boldsymbol{\mathbf{\Psi}}_{1}^{r}\right)^{T}\left(\boldsymbol{\mathbf{\Psi}}_{1}^{r}\right)^{*}\left(\frac{\partial\boldsymbol{h}^{r}}{\partial\boldsymbol{\theta}^{s}}\right)^{H}\right\} +2\left(\sigma_{n}^{r}\right)^{-2}\mathrm{Re}\left\{ \frac{\partial\boldsymbol{h}^{r}}{\partial\boldsymbol{\theta}^{r}}\left(\boldsymbol{\mathbf{\Psi}}_{2}^{r}\right)^{T}\left(\boldsymbol{\mathbf{\Psi}}_{2}^{r}\right)^{*}\left(\frac{\partial\boldsymbol{h}^{r}}{\partial\boldsymbol{\theta}^{s}}\right)^{H}\right\} ,
\end{equation}
\begin{align}
\mathbf{J}\left(\boldsymbol{\theta}^{s},\boldsymbol{\theta}^{s}\right) & =2\left(\sigma_{n}^{r}\right)^{-2}\mathrm{Re}\left\{ \frac{\partial\boldsymbol{h}^{r}}{\partial\boldsymbol{\theta}^{s}}\left(\boldsymbol{\mathbf{\Psi}}_{1}^{r}\right)^{T}\left(\boldsymbol{\mathbf{\Psi}}_{1}^{r}\right)^{*}\left(\frac{\partial\boldsymbol{h}^{r}}{\partial\boldsymbol{\theta}^{s}}\right)^{H}\right\} +2\left(\sigma_{n}^{r}\right)^{-2}\mathrm{Re}\left\{ \frac{\partial\boldsymbol{h}^{r}}{\partial\boldsymbol{\theta}^{s}}\left(\boldsymbol{\mathbf{\Psi}}_{2}^{r}\right)^{T}\left(\boldsymbol{\mathbf{\Psi}}_{2}^{r}\right)^{*}\left(\frac{\partial\boldsymbol{h}^{r}}{\partial\boldsymbol{\theta}^{s}}\right)^{H}\right\} \nonumber \\
 & +2\left(\sigma_{n}^{c}\right)^{-2}\mathrm{Re}\left\{ \frac{\partial\boldsymbol{h}^{c}}{\partial\boldsymbol{\theta}^{s}}\left(\boldsymbol{\mathbf{\Psi}}^{c}\right)^{T}\left(\boldsymbol{\Psi}^{c}\right)^{*}\left(\frac{\partial\boldsymbol{h}^{c}}{\partial\boldsymbol{\theta}^{s}}\right)^{H}\right\} ,
\end{align}
\begin{equation}
\mathbf{J}\left(\boldsymbol{\theta}^{c},\boldsymbol{\theta}^{s}\right)=2\left(\sigma_{n}^{c}\right)^{-2}\mathrm{Re}\left\{ \frac{\partial\boldsymbol{h}^{c}}{\partial\boldsymbol{\theta}^{c}}\left(\boldsymbol{\mathbf{\Psi}}^{c}\right)^{T}\left(\boldsymbol{\Psi}^{c}\right)^{*}\left(\frac{\partial\boldsymbol{h}^{c}}{\partial\boldsymbol{\theta}^{s}}\right)^{H}\right\} .
\end{equation}
where $\boldsymbol{h}^{r}\triangleq vec\left[\left(\mathbf{H}^{r}\right)^{T}\right]$
and the aggregated pilot matrices $\boldsymbol{\mathbf{\Psi}}_{1}^{r},\boldsymbol{\mathbf{\Psi}}_{2}^{r}$
and $\boldsymbol{\Psi}^{c}$ are given by
\begin{equation}
\boldsymbol{\mathbf{\Psi}}_{1}^{r}=\left[\begin{array}{c}
\mathbf{I}_{M}\otimes\boldsymbol{v}_{1,1}^{T}\\
\ldots\\
\mathbf{I}_{M}\otimes\boldsymbol{v}_{1,P_{1}}^{T}
\end{array}\right],\boldsymbol{\mathbf{\Psi}}_{2}^{r}=\left[\begin{array}{c}
\mathbf{I}_{M}\otimes\boldsymbol{v}_{2,1}^{T}\\
\ldots\\
\mathbf{I}_{M}\otimes\boldsymbol{v}_{2,P_{2}}^{T}
\end{array}\right],\boldsymbol{\Psi}^{c}=\left[\begin{array}{c}
u_{1,1}\mathbf{\mathbf{I}}_{M}\\
\ldots\\
u_{1,Q}\mathbf{I}_{M}
\end{array}\right]
\end{equation}

In this section, we shall optimize the radar pilot (DP) in Stage 2
to refine the estimation performance of the AoAs $\boldsymbol{\theta}^{r},\boldsymbol{\theta}^{s}$
for radar targets. The Cramer-Rao (CR) matrix for $\boldsymbol{\theta}^{r},\boldsymbol{\theta}^{s}$
is given by
\begin{equation}
\textrm{\textbf{CRB}\ensuremath{\mathbf{\triangleq J}_{\textrm{eff}}^{-1}}},
\end{equation}
where $\mathbf{J}_{\textrm{eff}}$ denotes the Equivalent Fisher Information
Matrix (EFIM) of radar targets given by (\ref{eq:efim}). The diagonal
elements of the CR matrix provide a lower bound for the MSE of any
unbiased estimator of $\boldsymbol{\theta}^{r},\boldsymbol{\theta}^{s}$.

\begin{equation}
\mathbf{J}_{\textrm{eff}}=\left[\begin{array}{cc}
\mathbf{J}\left(\boldsymbol{\theta}^{r},\boldsymbol{\theta}^{r}\right) & \mathbf{J}\left(\boldsymbol{\theta}^{r},\boldsymbol{\theta}^{s}\right)\\
\mathbf{J}\left(\boldsymbol{\theta}^{r},\boldsymbol{\theta}^{s}\right)^{T} & \mathbf{J}\left(\boldsymbol{\theta}^{s},\boldsymbol{\theta}^{s}\right)-\mathbf{J}\left(\boldsymbol{\theta}^{s},\boldsymbol{\theta}^{c}\right)\mathbf{J}^{-1}\left(\boldsymbol{\theta}^{c},\boldsymbol{\theta}^{c}\right)\mathbf{J}^{T}\left(\boldsymbol{\theta}^{s},\boldsymbol{\theta}^{c}\right)
\end{array}\right]\label{eq:efim}
\end{equation}

\subsection{Problem Formulation for Pilot Design}

There are three commonly used criteria involving a scalar measure
of the CR matrix \cite{crb_eign}. The first criterion is associated
with the minimization of the log-determinant of the CR matrix corresponding
to the minimization of the volume of the confidence ellipsoid. The
second criterion is the minimization of the trace of the CR matrix,
which is associated with the sum of squared errors. The third criterion
is the minimization of the maximal eigenvalue, $\lambda_{\max}$,
of the CR matrix. This criterion is associated with minimizing the
worst-case (largest) squared error. In practice, the desired pilot
must guarantee the sensing performance of the worst target. Therefore,
we adopt the third criterion in this paper.

Since the minimization of the maximal eigenvalue, $\lambda_{\max}^{\text{CR}}$,
of the CR matrix is equivalent to maximizing of the minimal eigenvalue,
$\lambda_{\min}^{\text{EFIM}}$, of the EFIM $\mathbf{J}_{\textrm{eff}}$,
the optimization problem for DP (radar pilot) design can be formulated
as
\begin{align}
\mathcal{P}: & \underset{\left\{ \boldsymbol{v}_{2,p}\right\} _{p=1}^{P_{2}}}{\max}\lambda\nonumber \\
\text{s.t. } & \textrm{tr}\left(\boldsymbol{v}_{2,p}\boldsymbol{v}_{2,p}^{H}\right)\leq P_{t},\label{eq:epigrapg_pilot_design}\\
 & \mathbf{J}_{\textrm{eff}}\left(\left\{ \boldsymbol{v}_{2,p}\right\} _{p=1}^{P_{2}}\right)\succeq\lambda\mathbf{I},\nonumber 
\end{align}
where $P_{t}$ is the transmit power for target detection, and we
have explicitly written $\mathbf{J}_{\textrm{eff}}$ as a function
of the optimization variables. Note that the constraint $\mathbf{J}_{\textrm{eff}}\left(\left\{ \boldsymbol{v}_{2,p}\right\} _{p=1}^{P_{2}}\right)\succeq\lambda\mathbf{I}$
ensures that $\lambda_{\min}^{\text{EFIM}}\geq\lambda$. Therefore,
Problem $\mathcal{P}$ maximizes the minimal eigenvalue $\lambda_{\min}^{\text{EFIM}}$
of the EFIM $\mathbf{J}_{\textrm{eff}}$.

\subsection{Pilot Optimization Algorithm}

Notice that Problem $\mathcal{P}$ is not a convex optimization problem
since the constraint $\mathbf{J}_{\textrm{eff}}\left(\left\{ \boldsymbol{v}_{2,p}\right\} _{p=1}^{P_{2}}\right)\succeq\lambda\mathbf{I}$
is not convex w.r.t $\left\{ \boldsymbol{v}_{2,p}\right\} _{p=1}^{P_{2}}$.
It can be observed that each submatrix of the FIM can be rewritten
as a function of $\left\{ \mathbf{V}_{2,p}\triangleq\boldsymbol{v}_{2,p}\boldsymbol{v}_{2,p}^{H}\right\} _{p=1}^{P_{2}}$
as follows,

\begin{align}
\mathbf{J}\left(\boldsymbol{\theta}^{r},\boldsymbol{\theta}^{r}\right) & =2\left(\sigma_{n}^{r}\right)^{-2}\mathrm{Re}\left\{ \frac{\partial\boldsymbol{h}^{r}}{\partial\boldsymbol{\theta}^{r}}\left(\boldsymbol{\mathbf{\Psi}}_{2}^{r}\right)^{T}\left(\boldsymbol{\mathbf{\Psi}}_{2}^{r}\right)^{*}\left(\frac{\partial\boldsymbol{h}^{r}}{\partial\boldsymbol{\theta}^{r}}\right)^{H}\right\} +\textrm{constant}\nonumber \\
= & 2\left(\sigma_{n}^{r}\right)^{-2}\mathrm{Re}\left\{ \frac{\partial\boldsymbol{h}^{r}}{\partial\boldsymbol{\theta}^{r}}\textrm{BlockDiag}\left(\sum_{p}\mathbf{V}_{2,p},\cdots,\sum_{p}\mathbf{V}_{2,p}\right)\left(\frac{\partial\boldsymbol{h}^{r}}{\partial\boldsymbol{\theta}^{r}}\right)^{H}\right\} +\textrm{constant}.
\end{align}

Motivated by the above observation, we convert the original problem
into a semi-definite programming with rank-1 constraints by introducing
new variables $\left\{ \mathbf{V}_{2,p}=\boldsymbol{v}_{2,p}\boldsymbol{v}_{2,p}^{H}\right\} _{p=1}^{P_{2}}$.
The optimization problem can then be equivalently reformulated as

\begin{align}
\mathcal{P}_{1}: & \underset{\left\{ \mathbf{V}_{2,p}\right\} _{p=1}^{P_{2}}}{\max}\lambda\nonumber \\
\text{s.t. } & \textrm{tr}\left(\mathbf{V}_{2,p}\right)\leq P_{t},\label{eq:powcon}\\
 & \mathbf{J}_{\textrm{eff}}\left(\left\{ \mathbf{V}_{2,p}\right\} _{p=1}^{P_{2}}\right)\succeq\lambda\mathbf{I},\label{eq:Jeffcon}\\
 & \textrm{rank}\left(\mathbf{V}_{2,p}\right)=1,p=1,2\ldots,P_{2}.\nonumber 
\end{align}
It can be shown that $\mathbf{J}_{\textrm{eff}}\left(\left\{ \mathbf{V}_{2,p}\right\} _{p=1}^{P_{2}}\right)\succeq\lambda\mathbf{I}$
is a convex constraint. However, the rank-1 constraints $\textrm{rank}\left(\mathbf{V}_{2,p}\right)=1,p=1,2\ldots,P_{2}$
are still non-convex. To overcome this challenge, we propose to replace
the rank-1 constraint with a tight and smooth approximation as stated
in the following lemma.
\begin{lem}
\label{lem:rank-appro}The rank of a positive semi-definite matrix
$\mathbf{V}\in\mathbb{C}^{M\times M}$ satisfies
\begin{align}
\textrm{rank} & \left(\mathbf{V}\right)=\lim_{\varepsilon\rightarrow0}\frac{M\log\left(\frac{1}{\varepsilon}\right)+\log\left|\mathbf{V}+\varepsilon\mathbf{I}\right|}{\log\left(1+\frac{1}{\varepsilon}\right)}.\label{eq:rankV}
\end{align}
Moreover, for any given $\varepsilon>0$, the RHS of (\ref{eq:rankV})
is a concave function of $\mathbf{V}$.
\end{lem}
\begin{IEEEproof}
Please refer to Appendix \ref{subsec:Proof-of-Lemma} for the proof.
\end{IEEEproof}
Using Lemma \ref{lem:rank-appro}, $\mathcal{P}_{1}$ can be well
approximated by the following problem for small $\varepsilon$
\begin{align}
\mathcal{P}_{\varepsilon}: & \underset{\left\{ \mathbf{V}_{2,p}\right\} _{p=1}^{P_{2}}}{\max}\lambda\nonumber \\
\text{s.t.} & \text{(\ref{eq:powcon}) and (\ref{eq:Jeffcon}),}\nonumber \\
 & M\log\left(\frac{1}{\varepsilon}\right)+\log\left|\mathbf{V}_{2,p}+\varepsilon\mathbf{I}\right|\leq\log\left(1+\frac{1}{\varepsilon}\right),\forall p.\label{eq:pilot_opt_cvx-1}
\end{align}
In fact, it can be shown that the optimal solution of $\mathcal{P}_{\varepsilon}$
converges to that of $\mathcal{P}_{1}$ as $\varepsilon\rightarrow0$.
Since $M\log\left(\frac{1}{\varepsilon}\right)+\log\left|\mathbf{V}_{2,p}+\varepsilon\mathbf{I}\right|$
is concave, we can apply the MM method to find a stationary point
of $\mathcal{P}_{1}$. Specifically, the MM method starts from an
initial point $\left\{ \mathbf{V}_{2,p}^{0}\right\} _{p=1}^{P_{2}}$,
and in the $i$-th iteration, it solves a locally convex approximation
of $\mathcal{P}_{1}$ around $\left\{ \mathbf{V}_{2,p}^{i-1}\right\} _{p=1}^{P_{2}}$
to obtain the next iterate $\left\{ \mathbf{V}_{2,p}^{i}\right\} _{p=1}^{P_{2}}$
as:
\begin{align}
\mathcal{P}_{c}: & \underset{\left\{ \mathbf{V}_{2,p}\right\} _{p=1}^{P_{2}}}{\max}\lambda\nonumber \\
\text{s.t.} & \text{ (\ref{eq:powcon}) and (\ref{eq:Jeffcon}),}\label{eq:pilot_opt_cvx-1-1}\\
 & \textrm{tr\ensuremath{\left\{  \left[\mathbf{V}_{2,p}^{i-1}+\varepsilon\mathbf{I}\right]^{-1}\left[\mathbf{V}_{2,p}-\mathbf{V}_{2,p}^{i-1}\right]\right\} } }\nonumber \\
\leq & \log\left(1+\frac{1}{\varepsilon}\right)-M\log\left(\frac{1}{\varepsilon}\right)-\log\left|\mathbf{V}_{2,p}^{i-1}+\varepsilon\mathbf{I}\right|,\forall p,
\end{align}
where (\ref{eq:pilot_opt_cvx-1-1}) is obtained by the first-order
Taylor expansion of the constraint function in (\ref{eq:pilot_opt_cvx-1}).

The overall algorithm is summarized as in Algorithm \ref{alg:The-Pilot-Optimization}.
To ensure the rank-1 constraints are strictly satisfied, a rank-1
projection is adopted in the final step as $\boldsymbol{v}_{2,p}^{*}=\mathbf{\tilde{V}}_{2,p}^{*}\left(:,1\right)$,
where $\mathbf{\tilde{V}}_{2,p}^{*}\left(:,1\right)$ is the dominant
eigenvector of $\mathbf{V}_{2,p}^{*}$. Note that since the constraint
function in (\ref{eq:pilot_opt_cvx-1}) is a very good approximation
of the rank-1 constraint, $\mathbf{V}_{2,p}^{*}$ will be close to
a rank-1 matrix and thus the performance loss caused by the rank-1
projection is tiny. Algorithm \ref{alg:The-Pilot-Optimization} requires
the knowledge of AoAs $\boldsymbol{\theta}$ and radar/communication
channels $\boldsymbol{h}^{c},\boldsymbol{h}^{r}$, whose estimated
values can be obtained using the Turbo-SBI algorithm in Stage 1. In
the simulations, we show that the performance loss caused by using
the estimated values of $\boldsymbol{\theta}$ and $\boldsymbol{h}^{c},\boldsymbol{h}^{r}$
in Stage 1 is acceptable.

\begin{algorithm}[tbh]
{\small{}\caption{The Pilot Optimization Algorithm\label{alg:The-Pilot-Optimization}}
}{\small\par}

\textbf{Input:} $P_{t}$, \textcolor{black}{a feasible $\mathbf{V}_{2,p}^{0}$,
maximum iteration numbers $I_{max}$, threshold $\epsilon$.}

\textbf{Output:} $\boldsymbol{v}_{2,p}^{*}$,$\forall p$.

\begin{algorithmic}

\FOR{${\color{blue}{\color{black}i=1,\cdots,I_{max}}}$}

\STATE Obtain the next iterate $\left(\mathbf{V}_{2,p}\right)^{i}$
by solving the locally convex approximation problem $\mathcal{P}_{c}$.

\IF{\textcolor{black}{$\left\Vert \mathbf{J}_{\textrm{eff}}\left(\left\{ \mathbf{V}_{2,p}^{i}\right\} _{p=1}^{P_{2}}\right)-\mathbf{J}_{\textrm{eff}}\left(\left\{ \mathbf{V}_{2,p}^{i-1}\right\} _{p=1}^{P_{2}}\right)\right\Vert \leq\epsilon$}}

\STATE \textbf{\textcolor{black}{break}}

\ENDIF

\ENDFOR

\STATE Let $\boldsymbol{v}_{2,p}^{*}=\mathbf{\tilde{V}}_{2,p}^{*}\left(:,1\right)$,$\forall p$
and output $\mathbf{V}_{2,p}^{*}$,$\forall p$.

\end{algorithmic}
\end{algorithm}

\section{Simulation Results}

In this section, we shall use simulations under the CDL channel model
in 3GPP R15 \cite{3gpp_Rel15} to verify that the proposed J-PoTdCe
scheme can achieve superior performance over the following baseline
schemes/algorithms.
\begin{itemize}
\item \textbf{Separate design using ML-based two-step detection and estimation
(SD-MLTS) \cite{Caire_OTFSsensing}:} The target detection and channel
estimation are performed separately using the ML-based two-step detection
and estimation algorithm in \cite{Caire_OTFSsensing} with omidirectional
pilots.
\item \textbf{Separate design using Turbo-SBI (SD-SBI): }The target detection
and channel estimation are performed separately using the proposed
Turbo-SBI algorithm (i.e., assuming independent sparse channel priors
for communication and radar sensing), with omidirectional pilots.
\item \textbf{Joint design with random pilots (JDRP):} The target detection
and channel estimation are performed jointly using the proposed Turbo-SBI
algorithm with omidirectional pilots.
\item \textbf{Joint design with SDR-based pilots (JDSDR):} The target detection
and channel estimation are performed jointly using the proposed Turbo-SBI
algorithm with the pilots optimized using the SDR method.
\item \textbf{Genie-aided J-PoTdCe:} This is the proposed scheme with the
pilots optimized based on the genie-aided information, i.e., the true
values of AoAs $\boldsymbol{\theta}$ and channels $\boldsymbol{h}^{c},\boldsymbol{h}^{r}$.
\end{itemize}
In the simulation, the BS is equipped with a ULA of $M=64$ antennas.
We set $P_{1}=P_{2}=1$. For convenience, define the common sparsity
ratio as $\rho_{c}=\left|\Omega_{r}\cap\Omega_{c}\right|/\Omega_{s}$.
When $\rho_{c}=1$, the radar and communication channels share the
same common AoA set $\Omega_{S}$. Therefore, $\rho_{c}$ reflects
the correlation among the AoA sets of the two channels. To control
the common sparsity ratio $\rho_{c}$ in the simulations, we first
generate the communication channel according to the CDL model. Then
we randomly choose a proper number of the AoAs of the communication
channel as part of the AoAs of the radar channel and the other AoAs
of the radar channel are generated similar to the CDL model. The received
SNR is set as $3\textrm{dB}$. For estimation performance, we compare
the average and worst-case MSE of the target AoA and the normalized
MSE (NMSE) of the communication and radar channels. We also compare
the target detection performance in terms of both false alarm probability
and miss detection probability. Specifically, for SD-MLTS, the target
detection method is given in (20) in \cite{Caire_OTFSsensing}. For
all other algorithms, the BS claims a target is detected around the
$m$-th AoA direction if $p^{post}(s_{m}^{r}=1)>0.5$, where $p^{post}(s_{m}^{r}=1)$
is the posterior probability of $s_{m}^{r}=1$ obtained by the detection
algorithm.

\subsection{Convergence Performance of Pilot Optimization Algorithm}

\begin{figure}[h]
\centering{}\textcolor{black}{}%
\begin{minipage}[t]{0.45\textwidth}%
\begin{center}
\textcolor{black}{\includegraphics[clip,width=80mm]{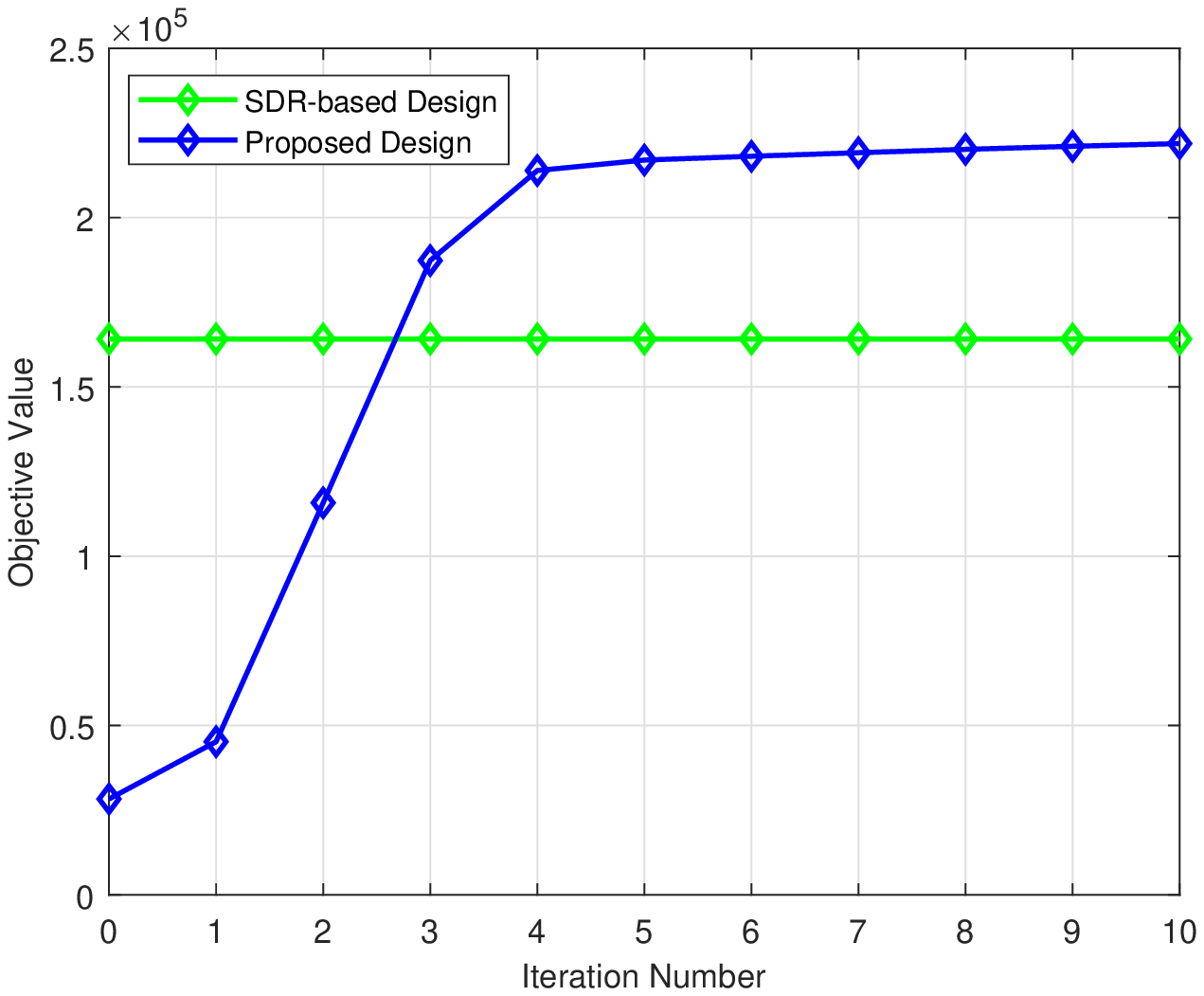}}\vspace{-14mm}
\par\end{center}
\textcolor{black}{\caption{\label{fig:converge-curve}The convergence of the pilot optimization
algorithm.}
}%
\end{minipage}\textcolor{black}{\hfill{}}%
\begin{minipage}[t]{0.45\textwidth}%
\begin{center}
\textcolor{black}{\includegraphics[clip,width=80mm]{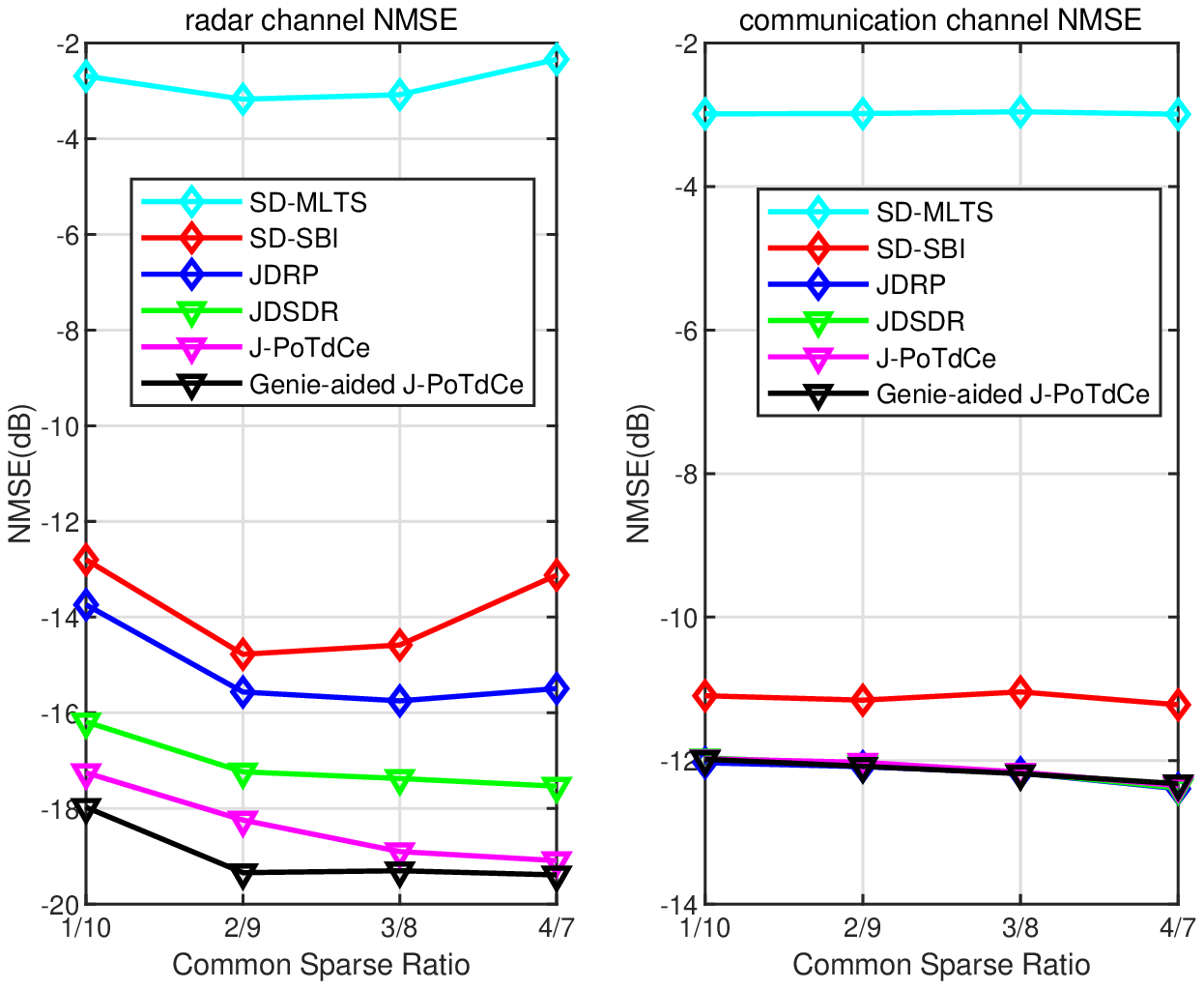}}
\par\end{center}
\vspace{-14mm}

\textcolor{black}{\caption{\textcolor{blue}{\label{fig:CENMSE}}Channel estimation NMSE versus
common sparsity ratio $\rho_{c}$.}
}%
\end{minipage}
\end{figure}

Fig. \ref{fig:converge-curve} illustrates the convergence of the
MM-based pilot optimization algorithm. As can be observed, the proposed
pilot optimization algorithm converges quickly within about 5 iterations,
and the achieved objective value is better than that of the SDR-based
pilot optimization algorithm.

\textcolor{black}{}

\textcolor{black}{}
\begin{figure}[h]
\centering{}\textcolor{black}{}%
\begin{minipage}[t]{0.45\textwidth}%
\begin{center}
\textcolor{black}{\includegraphics[clip,width=80mm]{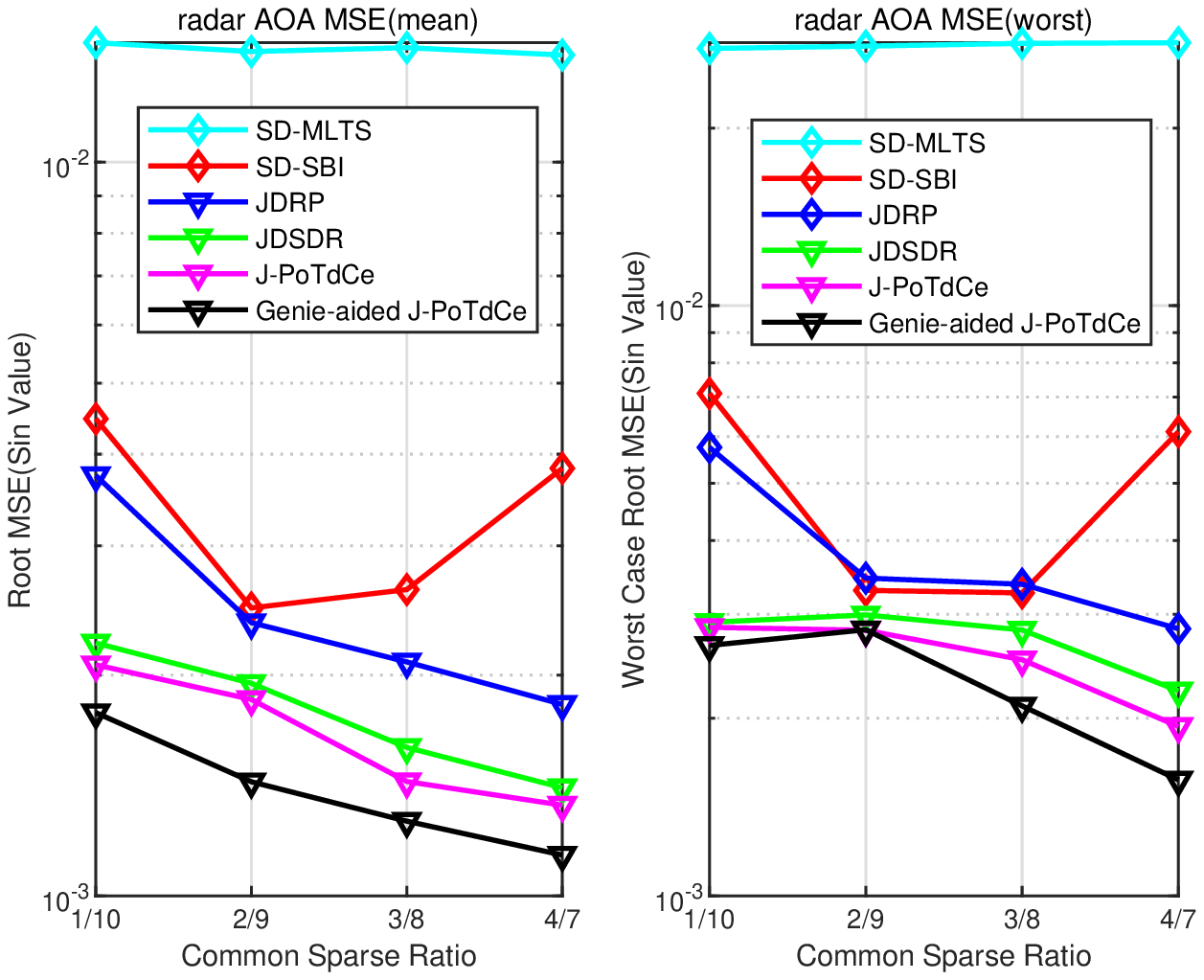}}\vspace{-14mm}
\par\end{center}
\textcolor{black}{\caption{\label{fig:AOAMSE}Average and worst-case MSE of the target AoA versus
common sparsity ratio $\rho_{c}.$}
}%
\end{minipage}\textcolor{black}{\hfill{}}%
\begin{minipage}[t]{0.45\textwidth}%
\begin{center}
\textcolor{black}{\includegraphics[clip,width=80mm]{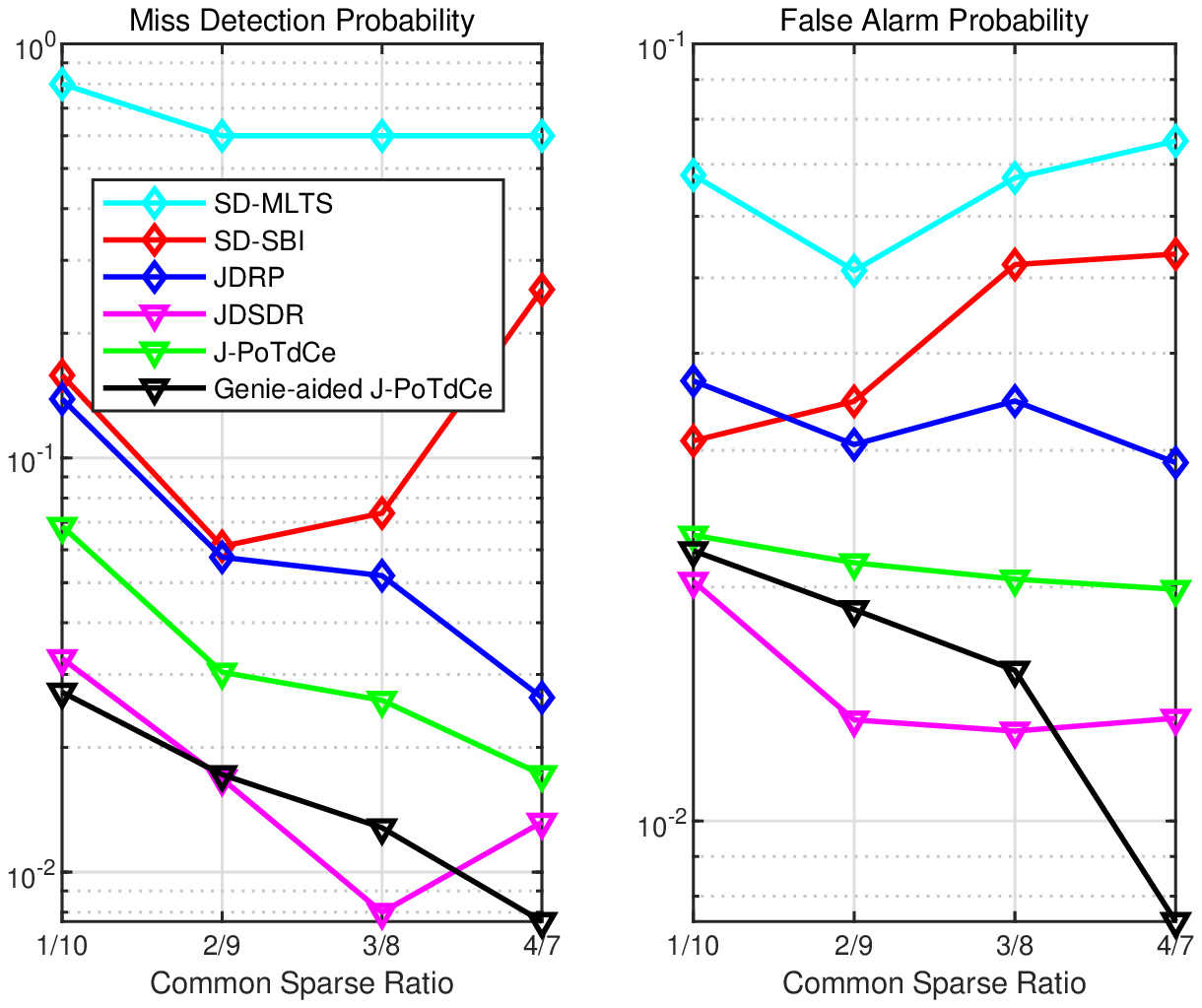}}
\par\end{center}
\vspace{-14mm}

\textcolor{black}{\caption{\label{fig:False-alarm-probability}False alarm probability and miss
detection probability versus common sparsity ratio $\rho_{c}.$}
}%
\end{minipage}
\end{figure}

\subsection{Impact of Common Sparsity Ratio}

In Fig. \ref{fig:CENMSE} - \ref{fig:False-alarm-probability}, we
compare the parameter estimation and target detection performance
versus the common sparsity ratio $\rho_{c}$, respectively. It can
be seen that the joint design achieves a better overall performance
than the separate design. Moreover, as $\rho_{c}$ increases, the
performance gap between the joint design and separate design increases
in general. This shows that the joint design approach can exploit
the joint sparsity between the radar and communication channels to
enhance the estimation/detection performance. Note that however, the
radar estimation/detection performance itself does not necessarily
improve with $\rho_{c}$ because the statistics (AoAs) of the radar
channel also changes with $\rho_{c}$. The proposed J-PoTdCe can achieve
a better performance than all practical baseline schemes (i.e., excluding
the genie-aided J-PoTdCe) for any given $\rho_{c}$, due to the exploitation
of the joint burst sparsity as well as the optimization of pilot.
Note that the performance gap between the proposed J-PoTdCe and gene-aided
J-PoTdCe is small, which verifies the feasibility of pilot optimization
based on the estimated information in Stage 1.

\section{Conclusion}

We proposed a two-stage joint pilot optimization, target detection
and channel estimation scheme to exploit the pilot beamforming gain
and joint burst sparsity of ISAC channels for enhanced target detection
and channel estimation performance. In Stage 1, the BS performs joint
target detection and channel estimation based on the reflected omidirectional
DP and received UP signals. In Stage 2, the BS exploits the prior
information obtained in Stage 1 to optimize the DP signal to achieve
beamforming gain and further refine the detection/estimation performance.
Specifically, a Turbo-SBI algorithm, which is a generalization of
the Turbo-OAMP in \cite{downlink_channel_estimation} from the PO
measurement matrix to arbitrary measurement matrix with dynamic grid
parameters, has been proposed for joint target detection and channel
estimation in both stages. The pilot optimization problem in Stage
2 is formulated as a worst-case CRB minimization problem, which contains
non-smooth rank-1 constraints. By replacing each rank-1 constraint
with a tight and smooth approximation, we developed an efficient pilot
optimization algorithm based on the MM method. Simulations verified
that the proposed scheme can achieve significant gain over baseline
schemes.

\appendix

\subsection{Message Update Equations for Module B of Turbo-SBI\label{subsec:Message-Update-Equations}}

\textbf{1) Message Passing Over the Path $x_{m}^{r}\rightarrow f_{m}^{r}\rightarrow s_{m}^{r}\rightarrow\eta_{m}^{r}\rightarrow s_{m}$:}

The message from variable node $x_{m}^{r}$ to factor node $f_{m}^{r}$
is
\begin{equation}
\nu_{x_{m}^{r}\rightarrow f_{m}^{r}}(x_{m}^{r})=\mathcal{CN}(x_{m}^{r};x_{B,m}^{r,pri},v_{B,m}^{r,pri}).\label{eq:message1}
\end{equation}
The message from factor node $f_{m}^{r}$ to variable node $s_{m}^{r}$
is
\begin{align}
\nu_{f_{m}^{r}\rightarrow s_{m}^{r}}(s_{m}^{r})\propto & \int\nu_{x_{m}^{r}\rightarrow f_{m}^{r}}(x_{m}^{r})\times f_{m}^{r}(x_{m}^{r},s_{m}^{r})dx_{m}^{r}\nonumber \\
\varpropto & \pi_{s^{r},m}^{in}\delta(s_{m}^{r}-1)+(1-\pi_{s^{r},m}^{in})\delta(s_{m}^{r}),\label{eq:message2}
\end{align}
where $\pi_{s^{r},m}^{in}=(1+\frac{\mathcal{CN}(0;x_{B,m}^{r,pri},v_{B,m}^{r,pri})}{\mathcal{CN}(0;x_{B,m}^{r,pri},v_{B,m}^{r,pri}+(\sigma_{m}^{r})^{2})})^{-1}$.
Then the message from variable node $s_{m}^{r}$ to factor node $\eta_{m}^{r}$
is the same as $\nu_{f_{m}^{r}\rightarrow s_{m}^{r}}(s_{m}^{r}).$
The message from factor node $\eta_{m}^{r}$ to variable node $s_{m}$
is
\begin{align}
\nu_{\eta_{m}^{r}\rightarrow s_{m}}(s_{m})= & \underset{\boldsymbol{s^{r}}}{\sum}\eta_{m}^{r}(s_{m}^{r},s_{m})\times\nu_{s_{m}^{r}\rightarrow\eta_{m}^{r}}(s_{m}^{r})\nonumber \\
= & \pi_{s,m}^{r,in}\delta(s_{m}-1)+(1-\pi_{s,m}^{r,in})\delta(s_{m}),\label{eq:message3}
\end{align}
where $\pi_{s,m}^{r,in}=(1+\frac{1-\pi_{s^{r},m}^{in}}{1+2\pi_{s^{r},m}^{in}\rho^{r}-\pi_{s^{r},m}^{in}-\rho^{r}})$$^{-1}$.

\textbf{2) The message passing over the path $x_{m}^{c}\rightarrow f_{m}^{c}\rightarrow s_{m}^{c}\rightarrow\eta_{m}^{c}\rightarrow s_{m}$}
is similar to that in 1) and thus is omitted for conciseness. The
final result is given by
\begin{equation}
\nu_{\eta_{m}^{c}\rightarrow s_{m}}(s_{m})=\pi_{s,m}^{c,in}\delta(s_{m}-1)+(1-\pi_{s,m}^{c,in})\delta(s_{m}),\label{eq:message4}
\end{equation}
where $\pi_{s,m}^{c,in}=(1+\frac{1-\pi_{s^{c},m}^{in}}{1+2\pi_{s^{c},m}^{in}\rho^{c}-\pi_{s^{c},m}^{in}-\rho^{c}})$$^{-1}$
and $\pi_{s^{c},m}^{in}=(1+\frac{\mathcal{CN}(0;x_{B,m}^{c,pri},v_{B,m}^{c,pri})}{\mathcal{CN}(0;x_{B,m}^{c,pri},v_{B,m}^{c,pri}+(\sigma_{m}^{c})^{2})})^{-1}$.

\textbf{3) Message Passing Over the Markov Chain of $\boldsymbol{s}$:}
\begin{equation}
\nu_{h_{m}\rightarrow s_{m}}(s_{m})\propto\gamma_{m}^{f}s_{m}+(1-\gamma_{m}^{f})(1-s_{m}),\label{eq:message5}
\end{equation}

\begin{equation}
\nu_{h_{m+1}\rightarrow s_{m}}(s_{m})\propto\gamma_{m}^{b}s_{m}+(1-\gamma_{m}^{b})(1-s_{m}),\label{eq:message6}
\end{equation}
where
\begin{equation}
\gamma_{m}^{f}=\frac{\rho_{0,1}(1-\pi_{m-1}^{in})(1-\gamma_{m-1}^{f})+\rho_{1,1}\pi_{m-1}^{in}\gamma_{m-1}^{f}}{(1-\pi_{m-1}^{in})(1-\gamma_{m-1}^{f})+\pi_{m-1}^{in}\gamma_{m-1}^{f}},\label{eq:message7}
\end{equation}

\begin{equation}
\gamma_{m}^{b}=\frac{\rho_{1,0}((\pi_{R,m}^{in})^{-1}-1)((\gamma_{m+1}^{b})^{-1}-1)+(1-\rho_{1,0})}{(\rho_{0,0}+\rho_{1,0})((\pi_{m+1}^{in})^{-1}-1)((\gamma_{m+1}^{b})^{-1}-1)+\rho_{1,1}+\rho_{0\text{,}1}},\label{eq:message8}
\end{equation}
with $\gamma_{1}^{f}=$$\frac{\rho_{0,1}}{\rho_{0,1}+\rho_{1,0}},\gamma_{M}^{b}=\frac{1}{2}$
and $\pi_{m}^{in}=\frac{\pi_{s,m}^{r,in}\pi_{s,m}^{c,in}}{\pi_{s,m}^{r,in}\pi_{s,m}^{c,in}+(1-\pi_{s,m}^{r,in})(1-\pi_{s,m}^{c,in})}.$

\textbf{4) Message Passing Over the Path $s_{m}\rightarrow\eta_{m}^{r}\rightarrow s_{m}^{r}\rightarrow f_{m}^{r}\rightarrow x_{m}^{r}$:}

The message from variable node $s_{m}$ to factor node $\eta_{m}^{r}$
is

\begin{align}
\nu_{s_{m}\rightarrow\eta_{m}^{r}}(s_{m})\propto & \nu_{h_{m}\rightarrow s_{m}}(s_{m})\times\nu_{h_{m+1}\rightarrow s_{m}}(s_{m})\times\nu_{\eta_{m}^{c}\rightarrow s_{m}}(s_{m})\nonumber \\
= & \pi_{s,m}^{r,out}\delta(s_{m}-1)+(1-\pi_{s,m}^{r,out})\delta(s_{m}),\label{eq:message9}
\end{align}
where $\pi_{s,m}^{r,out}=\frac{\gamma_{m}^{f}\gamma_{m}^{b}}{\gamma_{m}^{f}\gamma_{m}^{b}+((\pi_{s,m}^{c,in})^{-1}-1)(1-\gamma_{m}^{f})(1-\gamma_{m}^{b})}$.
The message from factor node $\eta_{m}^{r}$ to variable node $s_{m}^{r}$
is
\begin{align}
\nu_{\eta_{m}^{r}\rightarrow s_{m}^{r}}(s_{m}^{r})\propto & \underset{s_{m}}{\sum}\eta_{m}^{r}(s_{m}^{r},s_{m})\times\nu_{s_{m}\rightarrow\eta_{m}^{r}}(s_{m})\nonumber \\
= & \pi_{s^{r},m}^{out}\delta(s_{m}^{r}-1)+(1-\pi_{s^{r},m}^{out})\delta(s_{m}^{r}),\label{eq:message10}
\end{align}
where $\pi_{s^{r},m}^{out}=\pi_{s,m}^{r,out}\rho^{r}.$The message
from variable node $s_{m}^{r}$ to the factor node $f_{m}^{r}$ is
the same as $\nu_{\eta_{m}^{r}\rightarrow s_{m}^{r}}(s_{m}^{r}).$
The message from variable node $f_{m}^{r}$ to factor node $x_{m}^{r}$
is
\begin{align}
\nu_{f_{m}^{r}\rightarrow x_{m}^{r}}(s_{m}^{r})= & \underset{s_{m}^{r}}{\sum}\nu_{\eta_{m}^{r}\rightarrow s_{m}^{r}}(s_{m}^{r})\times f_{m}^{r}(x_{m}^{r},s_{m}^{r})\nonumber \\
= & \pi_{s^{r},m}^{out}\mathcal{CN}(x_{m}^{r};0,(\sigma_{m}^{r})^{2})+(1-\pi_{s^{r},m}^{out})\delta(x_{m}^{r}).\label{eq:message11}
\end{align}

\textbf{5) The message passing over $s_{m}\rightarrow\eta_{m}^{c}\rightarrow s_{m}^{c}\rightarrow f_{m}^{c}\rightarrow x_{m}^{c}$
is similar to that in 4).}

\subsection{Proof of Lemma \ref{lem:rank-appro} \label{subsec:Proof-of-Lemma}}

The rank of a positive semi-definite matrix $\mathbf{V\in\mathbb{C}}^{M\times M}$
is given by
\begin{equation}
\textrm{rank}\left(\mathbf{V}\right)=\sum_{m}u\left(\lambda_{m}\right)
\end{equation}
where $\left\{ \lambda_{m}\right\} _{m=1}^{M}$ denote the eigenvalues
of $\mathbf{V}$ and $u\left(x\right)$ denotes the step-function.
It is easy to see that
\begin{equation}
u\left(\lambda_{m}\right)=\frac{\log\left(1+\frac{\lambda_{m}}{\varepsilon}\right)}{\log\left(1+\frac{1}{\varepsilon}\right)}+o\left(\varepsilon\right),
\end{equation}
where $\lim_{\varepsilon\rightarrow0}o\left(\varepsilon\right)=0$.
Therefore, we have
\begin{align}
\textrm{rank}\left(\mathbf{V}\right) & =\sum_{m}\frac{\log\left(1+\frac{\lambda_{m}}{\varepsilon}\right)}{\log\left(1+\frac{1}{\varepsilon}\right)}+o\left(\varepsilon\right)\nonumber \\
 & =\frac{M\log\left(\frac{1}{\varepsilon}\right)+\log\left(\left|\mathbf{V}+\varepsilon\mathbf{I}\right|\right)}{\log\left(1+\frac{1}{\varepsilon}\right)}+o\left(\varepsilon\right)
\end{align}

Moreover, for any given $\varepsilon>0$, the RHS of (\ref{eq:rankV})
is a concave function of $\mathbf{V}$ since $\log\left|\mathbf{V}+\varepsilon\mathbf{I}\right|$
is a concave function of $\mathbf{V}$.


\end{document}